\documentclass [12pt,preprint]{aastex}

\usepackage{epsfig}

\begin{document}
\voffset-1cm
\newcommand{\gsim}{\hbox{\rlap{$^>$}$_\sim$}}

\title{The superluminal motion of Gamma-Ray-Burst sources\\
                and the complex afterglow of GRB 030329}

\author{Shlomo Dado\altaffilmark{1}, Arnon Dar\altaffilmark{1} and
A. De R\'ujula\altaffilmark{1,}\altaffilmark{2}}

\altaffiltext{1}{dado@phep3.technion.ac.il, arnon@physics.technion.ac.il, 
dar@cern.ch.\\
Physics Department and Space Research Institute, Technion, Haifa 32000, 
Israel}
\altaffiltext{2}{alvaro.derujula@cern.ch.
Theory Division, CERN, CH-1211 Geneva 23, Switzerland}


\begin{abstract}

The source of the very bright Gamma-Ray Burst GRB 030329 is close enough
to us for there to be a hope to measure or significantly constrain its
putative superluminal motion. Such a phenomenon is expected in the
``Cannonball'' (CB) model of GRBs. Recent  precise data on the optical
and radio afterglow of this GRB ---which demonstrated its very complex
structure--- allow us to pin down the CB-model's
prediction for the afterglow-source position as a function of time. It has
been stated that (the unpublished part of) the new radio data
``unequivocably disprove'' the CB model. We show how greatly exaggerated
that obituary announcement was, and how precise a refined analysis of the
data would have to be, to be still of interest.

\end{abstract}


\section{Introduction and summary}

The currently best-studied theories of Gamma-Ray Bursts (GRBs) and their
afterglows (AGs) are the {\it Fireball} models (see, e.g., Zhang \&
Meszaros 2003 for a recent review) and the {\it Cannonball} (CB) model
(see, e.g., Dar \& De R\'ujula 2003a; Dado, Dar \& De R\'ujula, 2002a;
2003a and references therein). The first set of models is often considered
to be {\it the standard model} of GRBs. In spite of their
similarly-sounding names, these two models are (or were initially)
completely different in
their basic hypothesis, in their description of the data, and in their
predictions. In this note we concentrate on a CB-model prediction which is
not (to date) a standard-model one, the apparently superluminal motion of
the source of GRBs and their afterglows: the ``cannonballs". In quite
exceptional cases ---relatively close-by GRBs with sufficiently bright
(radio) AGs--- it may be possible to observe this superluminal motion of
CBs relative to ``fixed stars''. Prior to GRB 030329, the case in which
a possible superluminal motion (Dar \& De R\'ujula 2000) came closest
to being observable was that of GRB 980425, which could ``almost'' be 
classified (Dar \& De R\'ujula 2003a) as an X-ray flash (XRF).
For XRFs, the observation of a superluminal motion
may be simpler than for GRBs, for the source's apparent displacement  
in the sky is proportional to the (small) observer's viewing angle,
and we interpret XRFs as jetted GRBs viewed at larger angles 
(Dar \& De R\'ujula 2003a; Dado, Dar \& De R\'ujula 2003f). In a
different model, this interpretation of XRFs has also been advocated
by Yamazaki, Yonetoku \& Nakamura (2003).

The CB-model is a very explicit elaboration of the original proposal
by Shaviv and Dar (1995): that the $\gamma$-rays of a GRB would 
be generated by inverse Compton Scattering (ICS) of stellar light
by the electron constituents of transient narrow jets, 
emitted in stellar processes leading to gravitational collapse.

In the CB model, long-duration GRBs, as well as XRFs
(Dar \& De R\'ujula 2003a; Dado, Dar \& De R\'ujula 2003f)  are produced 
in the explosions of {\it ordinary} core-collapse SNe,
akin to SN1998bw\footnote{The X-ray and radio signals
of the SN1998bw/GRB980425 pair are, in the CB model,
attributed and well fit to the CB's AG, depriving the SN of
these ``peculiar'' emissions. The observed large velocity of the
SN's ejecta is attributed to their being observed exceptionally
close to the jet axis. Neither this SN, nor its GRB, were
exceptional (Dar \& Plaga 1999, Dar \& De R\'ujula 2000,
Dado et al.~2002a, 2003a).} (Dar \& De R\'ujula 2000),
the first SN to be observed in ``association'' with a GRB
(GRB 980425, Galama et al.~1998). Two opposite jets of CBs 
are emitted in the process, travelling with initially
large Lorentz factors: $\gamma_0\sim 10^3$. The CBs
initially expand (in their rest system) at a velocity comparable to,
or smaller than, the speed of sound in a relativistic plasma
($c/\sqrt{3}$), so that the jet opening angle (subtended by
a CB's radius as observed from its emission point\footnote{We
are neglecting the initial CB's radius, presumably comparable
or not much bigger than that of the collapsed core of the parent star,
and thus entirely negligible by the time the GRB is emitted.}) is 
$\alpha_j<\! 1/(\gamma_0\,\sqrt{3})$. The CBs' highly relativistic motion 
collimates their emitted radiation ---the GRB and its AG--- within a forward
beam of characteristic opening angle $1/\gamma$.
An observer sees the ``Doppler-favoured'' jet, travelling
at a small angle $\theta={\cal{O}}(1/\gamma)$ relative to
the line of sight.
Typically $\theta>\alpha_j$, so that the jet's
opening angle can be neglected and the observer's angle
is the {\it only} relevant one.

In the CB model, during the AG, the jet opening angle {\it diminishes}
with time, while the beaming angle 
$1/\gamma(t)$ increases, both evolutions being due to the
CB's interaction with the matter of the interstellar medium
(Dado, Dar \& De R\'ujula, 2002a). As a consequence, the
AG's source becomes increasingly ``pointlike'', and its motion
in the sky can in principle be followed. Since a CB's motion
is relativistic for days or months of (observer's) time, its
apparent displacement in the sky is {\it superluminal} 
(Courdec 1939, Rees 1967), as we argued in Dar \& De R\'ujula (2000).

The closest-by GRB observed so far ---GRB 980425, at a redshift
$z=0.0085$--- came very close to having an observable superluminal motion
(see Dado, Dar \& De R\'ujula 2003a for a detailed discussion). In a GCN
note (Dar \& De R\'ujula 2003b) we argued that this motion may be
observable in the next-closest GRB (030329, at $z=0.1685$).  In this note
we sharpen the predictions for this putative motion, given the current
availability of precise optical data at early times and times later than
the first $\sim 6$ days (Lipkin et al.~2003 and references therein), as
well as sparse X-ray data (Marshall \& Swank 2003;  Marshall, Markwardt \&
Swank 2003; Tiengo et al.~2003) and abundant radio data (Sheth et al.~2003; 
Berger et al.~2003; Pooley 2003; Kuno et al.~2004).

\begin{figure}[]
\epsfig{file=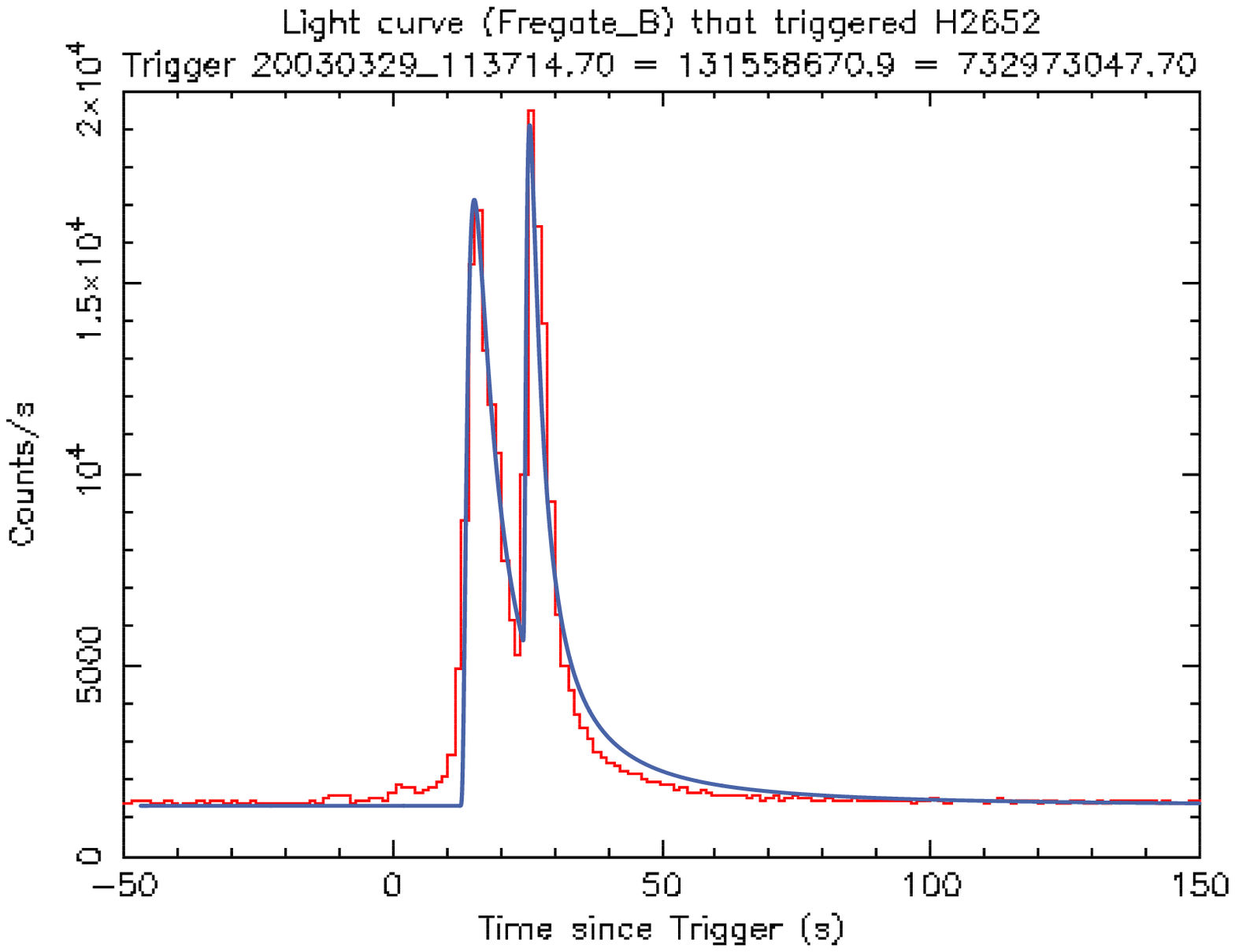, width=16cm}
\figcaption{The $\gamma$-ray light curve of GRB 030329, the (red)
binned curve (Vanderspeck et al.~2004); 
and its simplest CB-model description, the (blue) continuous
line (Dado et al.~2003c).
\label{GRB}} 
\end{figure}

A very relevant new information (Lipkin et al.~2003 and references therein) 
on the AG of GRB 030329 is the abundance of multiple deviations of the
optical light curves relative to a smoothly declining behaviour.  We shall
refer to these deviations as {\it features}. In Dado et al.~(2003c) we
attributed the most obvious optical-AG feature ---a ``shoulder" starting
at $t\sim$ 1 day (after burst)--- to a transition between a first to a
second dominant CB, a choice supported by the fact that the $\gamma$-ray
light-curve of this GRB has a very clear two-pulse structure (Vanderspek
et al.~2003; http:// space.mit.edu/ Hete/ Bursts/ GRB030329/; see also
Vanderspeck et al.~2004), as shown in Fig.~(\ref{GRB}). With the emergence
of a handful of other similarly-significant features in the first week
---fast ups and downs of the optical fluences, $F_\nu(t)$, by some 20 to
40\%--- a more elaborate interpretation is required. In the AG model
$F_\nu(t)$ is a direct and {\it quasi-local} tracer of the density of the
ISM through which a CB travels, a fact to which we
have attributed previous similar observations, e.g.~the ``humps" in the
optical AGs of GRBs 000301c and 970508 (Dado et al.~2002a).  It is
therefore necessary to investigate the effect of local ISM
density-inhomogeneities on the expected superluminal motion. This is what
we do in this note for the case of GRB 030329.
We conclude that the sources' motion results in a
displacement of $\sim 0.3$ (0.6) mas from day 3 to day 30 (100) after
burst. Detecting such a motion may not be out of the question.

For the benefit of readers not familiar with the CB model, we present in
an appendix a brief overview of the model and its current confrontation
with data. We also offer there some commentary on the evolution of the
standard model.

\section{The deceleration of a CB}

The Lorentz factor $\gamma=\gamma(x)$ of a CB diminishes with the distance
$x$ from the parent SN, as its motion is decelerated by collisions with
the constituents of the interstellar medium (ISM). 
In an approximately hydrogenic ISM of local  number density $n_p(x)$,
the evolution of $\gamma(x)$ is determined by energy-momentum conservation
to be:
 \begin{eqnarray}
{d\gamma\over \gamma^2}&=&-\,{dx\over L(x)}
\label{dgamma1}\\
L(x)&\equiv&
{N_{_{CB}}\over\pi\, R_\infty^2\,n_p(x)}\, ,
\label{dgamma2} 
\end{eqnarray}
with $N_{_{CB}}$ the CB's baryon number and $R_\infty$  
its calculable asymptotic radius, reached
within minutes of observer's time. 
For typical parameters, by the time $R(x)\approx R_\infty$, the function
$\gamma(x)$ is still extremely well approximated by its initial value
$\gamma_0\equiv \gamma(0)$ (Dado et al.~2002a).

For an object travelling at nearly the speed of light
the relation between $dx$ and the element of observer's time, $dt$, is:
\begin{equation}
{dx\over \gamma(x)\, \delta(x)}\simeq { c\,dt\over 1+z}\; ,
\label{dxdt}
\end{equation}
where $z$ is the cosmological redshift and $\delta(x)$
is the Doppler factor by which the energy of
photons emitted by the CB is locally boosted by its
relativistic motion.
For the Lorentz factors and viewing angles
relevant to our discussion,
$\gamma^2 \gg 1$ and $\theta^2 \ll 1$, and
\begin{equation}
  \delta = {1\over \gamma\, (1-\beta\, \cos\theta)}\approx
           {2\gamma\over 1+\gamma^2\, \theta^2}\; ,
\label{Doppler}
\end{equation}
to an excellent approximation.

We may integrate Eqs.~(\ref{dgamma1},\ref{dgamma2}) to obtain:
\begin{equation}
{1\over \gamma(x)}-{1\over \gamma_0}=\int_0^x {dx'\over L(x')}\, ,
\label{gammaint}
\end{equation}
which, for any given $n_p(x)$, yields the explicit $x$-dependent
Lorentz factor $\gamma(x)$. Substitute this result, with use
of Eq.~(\ref{Doppler}), into Eq.~(\ref{dxdt}) and integrate:
\begin{equation}
t= {1+z\over c}\, \int_0^x{dx\over \gamma(x)\,\delta(x)}\, ,
\label{tint}
\end{equation}  
to obtain the explicit relation between observer's time $t$ and travelled
distance $x\,.$
The last two equations are a parametric description of
the function of actual interest, $\gamma(t)$.

For some simple density profiles the above formal exercise can be carried
explicitly to the end. For a constant-density ISM, for instance, we may define:
\begin{equation}
L(x)=x_\infty\equiv{N_{_{B}}\over\pi\, R_\infty^2\, n_p}\, ,
\label{xinf}
\end{equation}
and conclude that $\gamma(t)$ is the real root of the cubic:
\begin{equation}
{1\over\gamma^3}-{1\over\gamma_0^3}
+3\,\theta^2\,\left[{1\over\gamma}-{1\over\gamma_0}\right]=
{2\,c\, t\over 3\, (1+z)\, x_\infty}\; ,
\label{cubic0}
\end{equation}
that is:
\begin{eqnarray}
 \gamma&=&\gamma(\gamma_0,\theta,x_\infty;t)
= {B^{-1}} \,\left[\theta^2+C\,\theta^4+{1/C}\right]\, ,\nonumber\\
 C&\equiv&
\left[{2/
\left(B^2+2\,\theta^6+B\,\sqrt{B^2+4\,\theta^6}\right)}\right]^{1/3}\, ,
\nonumber\\
 B&\equiv&
{1/ \gamma_0^3}+{3\,\theta^2/\gamma_0}+
{6\,c\, t/ [(1+z)\, x_\infty]}\, ,
\label{cubic}
\end{eqnarray}
while the distance travelled by a CB at a given observer's time is:
\begin{equation}
x(t)=x_\infty \,\left[{1\over \gamma(t)}-{1\over\gamma_0}\right]\; ,
\label{xoft}
\end{equation}
so that it takes a distance $x_{1/2}\equiv x_\infty/\gamma_0$ for
the CB to half its original Lorentz factor. As we shall see in detail
in the case of GRB 030329, by the time the distance $x_{1/2}$ is
reached, the AG fluence, proportional to a high power of $\gamma(t)$,
has decreased by more than one order of magnitude. 

We have performed CB-model fits to the AG measurements of all
the GRBs of known redshift, and found that the values of
$x_{1/2}$ are spread over more than an order of magnitude, centering
at $\sim 100$ pc (Dado et al 2002a). These are comparable
to the radii of the {\it superbubbles} wherein most SN explosions take
place,  created by the stellar winds and explosions of many previous  
SNe in dense star-formation  regions. The approximation of a constant
density $n_p$, employed in the quoted fits, may be a fair first try
for the density within the superbubble and in the galactic-halo regions
close-to ---but without--- the superbubble.  In the case of GRB 030329
we shall see that the density profile of the superbubble's surface,
as traced by the CB's X-ray and optical fluences, is quite rich 
and interesting.

\section{The superluminal motion of a CB}

The transverse projected velocity in the sky of a CB relative
to its parent SN can be obtained from Eq.~(\ref{Doppler}):
\begin{equation}
V_{_{CB}}=\sin\theta\;{dx\over dt}\simeq {\gamma\, \delta\, \theta\over (1+z)}\; c\, ,
\label{supervelocity}
\end{equation}
which, for typical parameters, is extremely superluminal.
The resulting angular separation at time $t$ is:
\begin{equation}
\Delta \alpha(t)={1\over D_A}\int_0^t\,V_{_{CB}}(t')\;dt'\rightarrow
\theta\,{x_\infty\over D_A}\left[{1\over\gamma(t)}-{1\over\gamma_0}\right]\; ,
\label{Deltaalpha}
\end{equation}
where $D_A=D_L/(1+z)^2$ is the angular distance
to the SN/CB system and $D_L$ is the luminosity distance
(we use throughout a cosmology with $\Omega=1$ and 
$\Omega_\Lambda=0.7$). The last expression in the r.h.s.~of
Eq.~(\ref{Deltaalpha}) is valid for a constant-density ISM.

At late time, when $[\gamma(t)\theta]^2\ll 1$, Eq.~(\ref{Doppler}) 
implies that $\delta(t)\approx 2\, \gamma(t)$,
while Eq.~(\ref{cubic}) implies that when $3\, [\gamma(t)\theta]^2\ll 
1$, $\gamma(t)$ approaches its asymptotic behaviour,  
$\gamma(t)\approx [2\,c\, t/ 3\, (1+z)\, x_\infty]^{-1/3}$.
The corresponding asymptotic behaviour of $\Delta \alpha$ is,
for a constant $n_p$:
\begin{equation}
\Delta \alpha(t)\sim {\theta \over D_A}\;
\left[{2\;x_\infty^2\;c\,t \over 3\,(1+z)}\right]^{1/3}\,,
\label{latealpha}
\end{equation}
which, in practice, is a fair approximation for $t$ larger than a 
few days. Since XRFs are GRBs seen at a relatively large $\theta$,
Eq.~(\ref{latealpha}) implies that the superluminal motion 
in the former systems is
relatively easier to observe. Moreover, the larger the value
of $\theta$, the smaller the values of $D_A$ favoured by
selection effects (detection thresholds). This also privileges XRFs over GRBs
as potential targets for a search of a superluminal CB motion.

\section{The afterglow of GRBs}

In the CB model, the AGs of GRBs and XRFs consist of three contributions,
from the CBs themselves, the concomitant SN, and the host galaxy:
\begin{equation}
F_{AG}=F_{CBs}+F_{SN}+F_{HG}\, .
\label{sum}
\end{equation}
The latter contribution is usually determined
by late-time observations, when the CB and SN contributions become
negligible, or from measurements with sufficient angular resolution
to tell apart $F_{CBs}+F_{SN}$ from $F_{HG}$.

Let the unattenuated energy flux
density of SN1998bw at redshift $z_{bw}=0.0085$ (Galama et al.~1998)   
be $F_{bw}[\nu,t]$. For a similar SN placed at a redshift $z$ 
(Dar 1999, Dar \& De R\'ujula 2000, Dado et al.~2002a):   
\begin{equation}
F_{SN}[\nu,t] = {1+z \over 1+z_{bw}}\;
{D_L^2(z_{bw})\over D_L^2(z)}\, A(\nu,z)\, F_{bw}[\nu',t']\, ,
\label{bw}
\end{equation}
where $A(\nu,z)$ is the attenuation along the line of sight,
$\nu'=\nu\, (1+z)/ (1+z_{bw})$,  and $t'=t\, (1+z_{bw})/(1+z)$.     
The simple ansatz that {\it all} long-duration GRBs would be
associated with SN1998bw-like SNe
(Dar \& Plaga 1999; Dar \& De R\'ujula 2000; Dado et al.~2002a) has 
proven to be unexpectedly precise and successful.  
 For the most precise test so far,
that of the GRB 030329/SN2003dh pair, see, e.g., Dado et al.~(2003c),
Stanek et al.~(2003), Matheson et al.~(2003) and Hjorth et al.~(2003).

The AG of the CBs is mainly due to synchrotron radiation from 
accelerated  electrons in the CB's
chaotic magnetic field. At optical and higher frequencies,
the AG has the approximate form 
(Dado et al.~2003a,c):
\begin{equation}
F_{_{CB}}[\nu,t]\propto n_p^{(1+\hat\alpha)/2}\, R_\infty^2\, 
\gamma^{3\hat\alpha-1}\,
\delta^{3+\hat\alpha}\, A(\nu,t)\, \nu^{-\hat\alpha}\, ,
\label{afterglow}
\end{equation}
with $\hat\alpha$ changing from $\sim 0.5$ to $\sim 1.1$
as each given observed frequency exceeds the time-dependent 
{\it ``injection bend'':}
\begin{equation}
 \nu_b(t) \simeq 1.87\times 10^3\, [\gamma(t)]^3\,
\left[{n_p\over 10^{-3}\;cm^3}\right]^{1/2}\, {\rm Hz}\, ,
\label{nubend}
\end{equation}
where $n_p$ is the baryon density of the interstellar 
medium\footnote{In the CB model, the spectral evolution as 
$\hat\alpha$ changes from $\sim\! 0.5$ to $\sim\! 1.1$,
is interpolated by $(\nu/\nu_b(t))^{-0.5}/
\sqrt{1+[\nu/\nu_b(t)]^{1.1}}$ (Dado et al.~2003a).}.  
In the same approximation in which Eq.~(\ref{latealpha}) was derived,
the AGs of GRBs and XRFs have the 
asymptotic behaviour $F_\nu\sim \nu^{-1.1\pm 0.1}\, t^{-2.13\pm 0.1}$
(Dado et al.~2002a). At radio frequencies, the AG spectrum
is affected by self-absorption in the CBs themselves,
characterizable by a single parameter per CB: a 
``free-free'' absorption frequency, $\nu_a$ (Dado et al.~2003a).

The attenuation $A(\nu,t)$ is a product of the attenuation in the host
galaxy, in the intergalactic medium, and in our Galaxy. The attenuation in
our galaxy in the direction of the GRB or XRF is usually estimated from the
Galactic maps of selective extinction, $E(B-V)$, of Schlegel, Finkbeiner \&
Davis (1998), using the extinction functions of Cardelli et al.~(1986). 
The extinction in the host galaxy and the intergalactic medium,
$ A(\nu,t)$
in Eq.~(\ref{bw}), can be estimated from the difference between the
observed spectral index {\it at very early time when the CBs are still
near the SN} and that expected in the absence of extinction. Indeed, the
CB model predicts ---and the data confirm with precision--- the gradual
evolution of the effective optical spectral index towards the constant
value $\approx -1.1$ observed in all ``late'' AGs (Dado et al.~2002a; 2003a). 
The ``late'' index is independent of the attenuation in the
host galaxy, since at $ t>1$ (observer's) days after the explosion, the
CBs are typically already moving in the low-column-density, 
optically-transparent halo of the host galaxy.


\section{Some lessons from past GRBs}

The CB-model description of the R-band AGs of six representative GRBs
of interest to the analysis of GRB 030329 are shown in Fig.~(\ref{tiriri}),
and discussed anon. 
In Dado et al.~(2002a) we showed that, in the CB model, the AGs of the 
GRBs of known redshift measured at the time 
could be successfully explained by a single dominant CB launched in
the explosion of a SN akin to SN1998bw and moving, after a couple of observer
hours, in an approximately constant-density ISM. A typical example is that of
GRB 990510. Two exceptions are 
GRB 990123 and GRB 021211, whose AGs were measured very early, at a time when
their CBs may  be piercing the expected $n_p\sim 1/r^2$ density profile
generated by the pre-SN ``wind'' of the parent star. In the CB
model, the early fluence of optical AGs is proportional
to $n_e^{3/4}\approx n_p^{3/4}\propto r^{-3/2}\approx t^{-3/2}$, in excellent
agreement with the observations (Dado et al.~2003b).

Prior to GRB 030329, there were three cases (GRBs 970508,  000301c and
021004) for which the data showed clear evidence for deviations from
a smooth AG behaviour. In the case of GRB 000301c, we
attributed the ``residua'' of the
observational data relative to its smooth CB-model fit to moderate
deviations of the ISM density (far from the parent SN) from a constant
value.  In the case of GRB 970508,
we have actually fit the AG to a jump from one to
another constant-density value (Dado et al.~2002a). 
For GRB 021004 we found it more natural to
fit the AG to the contribution of two CBs, since in this case the
$\gamma$-ray count-rate as a function of time has a very clear two-pulse
(i.e.~two-CB) structure (Dado et al.~2003d). Moreover, unlike for GRB   
970508, our attempts to describe the AG of GRB 021004 in terms of density
variations failed, while the two-CB description fits the general trend of
the AG very well.

GRB 030329 also has a two-CB $\gamma$-ray structure, but an
unprecedentedly non-smooth AG, presumably for the
simple reason that the precision and continuity of the data are
also unprecedented. In the case of this GRB, the question will arise 
whether or not we should interpret its peculiar AG shape to two
CBs, to density fluctuations, or to a combination of both effects.

\section{The afterglow of GRB 030329: the first two rounds}

In Dado et al.~(2003c) we made a ``first-round" CB-model fit to the 
NIR-optical observations 
of the AG of this GRB, then extending up to day $\sim 6$ after burst.
Since experience with all previously-measured AGs had given us
confidence in the quality of the model, we extrapolated these 
fits to later times, and
we predicted the presence of a SN akin to SN1998bw, luminous
enough to compete with the AG ---and be discovered---
ten days after burst\footnote{The fits discussed in the current paper would
have given the same result, since from day $\sim 6$ they have also
reached the predicted asymptotic behaviour $F_\nu\sim t^{-2.13}$, and
in the CB model there is no need to hypothesize whether ``breaks''
in the AG fluence have occurred by a given time, there being no breaks.}. 
SN2003dh was discovered 9.6 days after
burst   and turned out to be surprisingly similar
to SN1998bw (Stanek el al.~2003; Matheson et al.~2003; Hjorth et al.~2003).

\begin{figure}[]
\hskip 2truecm
\plotone{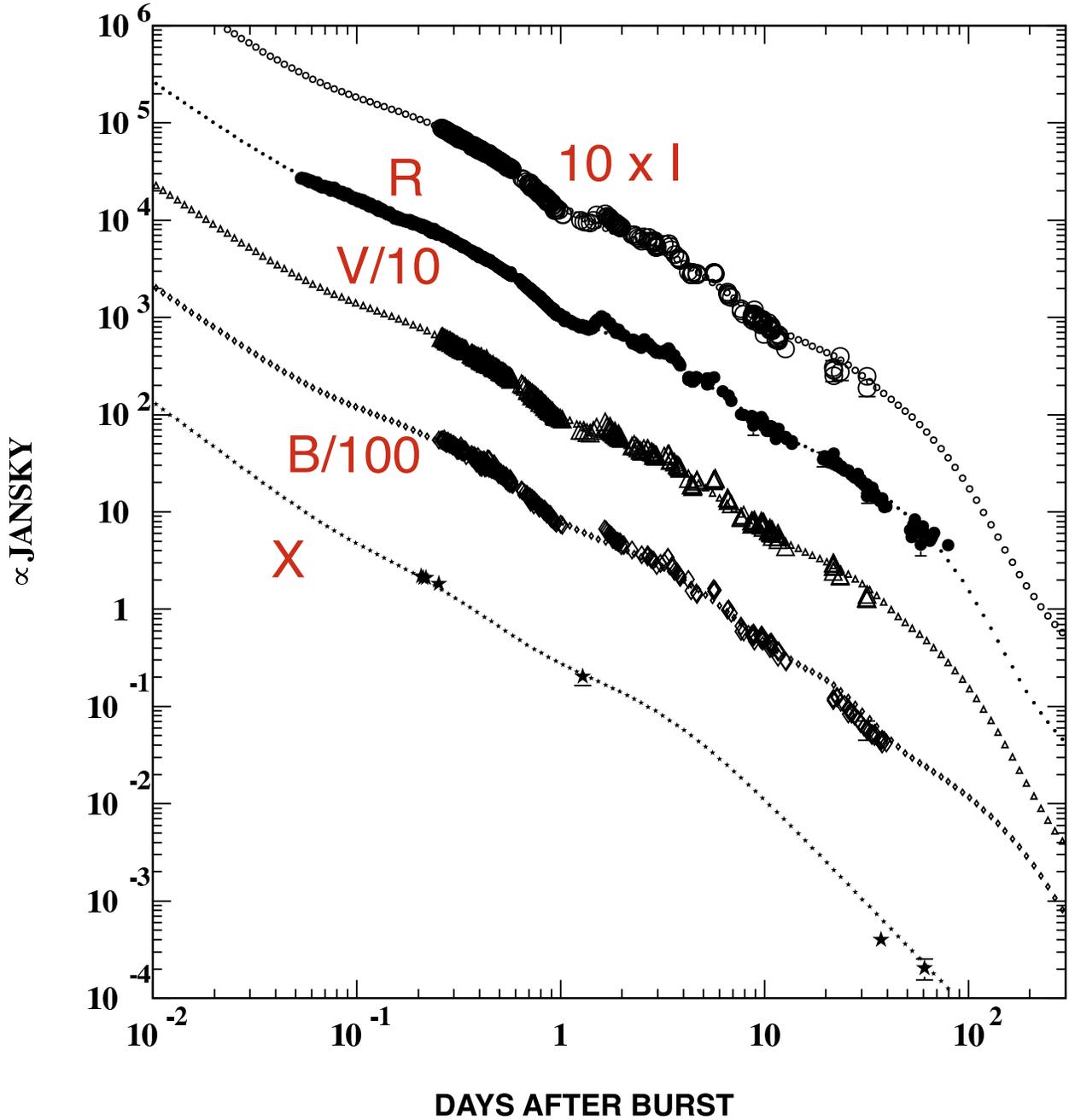}
\figcaption{The NIR--optical and X-ray observations of the AG of GRB 030329
and the ``second-round"
broad-band fit for two CBs with different parameters,  described 
in the text. The ISM density is assumed to be a 
constant plus a ``wind'' contribution decreasing as $ 1/r^2$.  The 
various bands are scaled for presentation. The fit is to
the X-ray data of RXTE (Marshall \& Swank, 2003;  Marshall, Markwardt \& 
Swank, 2003) and XMM-Newton (Tiengo et al.~2003) and many other NIR-optical 
measurements, recalibrated by Lipkin et al.~(2003 and references therein);
as well as the radio data of Sheth et al.~(2003) and Berger et al.~(2003),
which are shown in Fig.~(\ref{figthree}).
 The  host-galaxy's contribution
was neglected.  The individual bands have been rescaled for
clarity.
\label{figone}}
\end{figure}

In a second round, we have now extended our original fit to include the 
X-ray data of RXTE (Marshall \& Swank, 2003;  Marshall, Markwardt \& 
Swank, 2003) and XMM-Newton 
(Tiengo et al.~2003), as well as the radio data
of Sheth et al.~(2003) and Berger et al.~(2003)  and many more NIR-optical 
measurements (Lipkin et al.~2003 and references therein). The optical and X-ray 
data and the broad-band CB-model fit are shown in Fig.~(\ref{figone}).
The inclusion of the new data modifies the parameters of the
original data-poor fit only at the 10\% level, a satisfactory result.
The photon-number light-curve of the $\gamma$ rays of GRB 030329 
consisted of two clear pulses, corresponding in 
the CB model to two dominant CBs,  see Fig.~(\ref{GRB}). 
Consequently, the fit to the AG was  again
performed with the additive contribution of two separate CBs.
In the optical domain, the two contributions correspond to the
two shoulders, as illustrated in Fig.~(\ref{figtwo}), wherein we show the R-band 
results and the separate contributions of the two CBs.
The data variations relative to the predicted smooth AG light-curve, ups and
downs of $\sim 1/2$ magnitude, are, as for GRBs 970508, 000301c (Dado et
al.~2002a) and 021004 (Dado et al.~2003b), to be expected:  they 
trace moderate deviations from a constant-density interstellar medium,
as implied by Eqs.~(\ref{afterglow},\ref{nubend}). These density variations
are discussed in detail in the next section.

\begin{figure}[]
\hskip 2truecm
\vspace*{0.8cm}
\vspace*{-.4cm}
\hspace*{-1.4cm}
\plotone{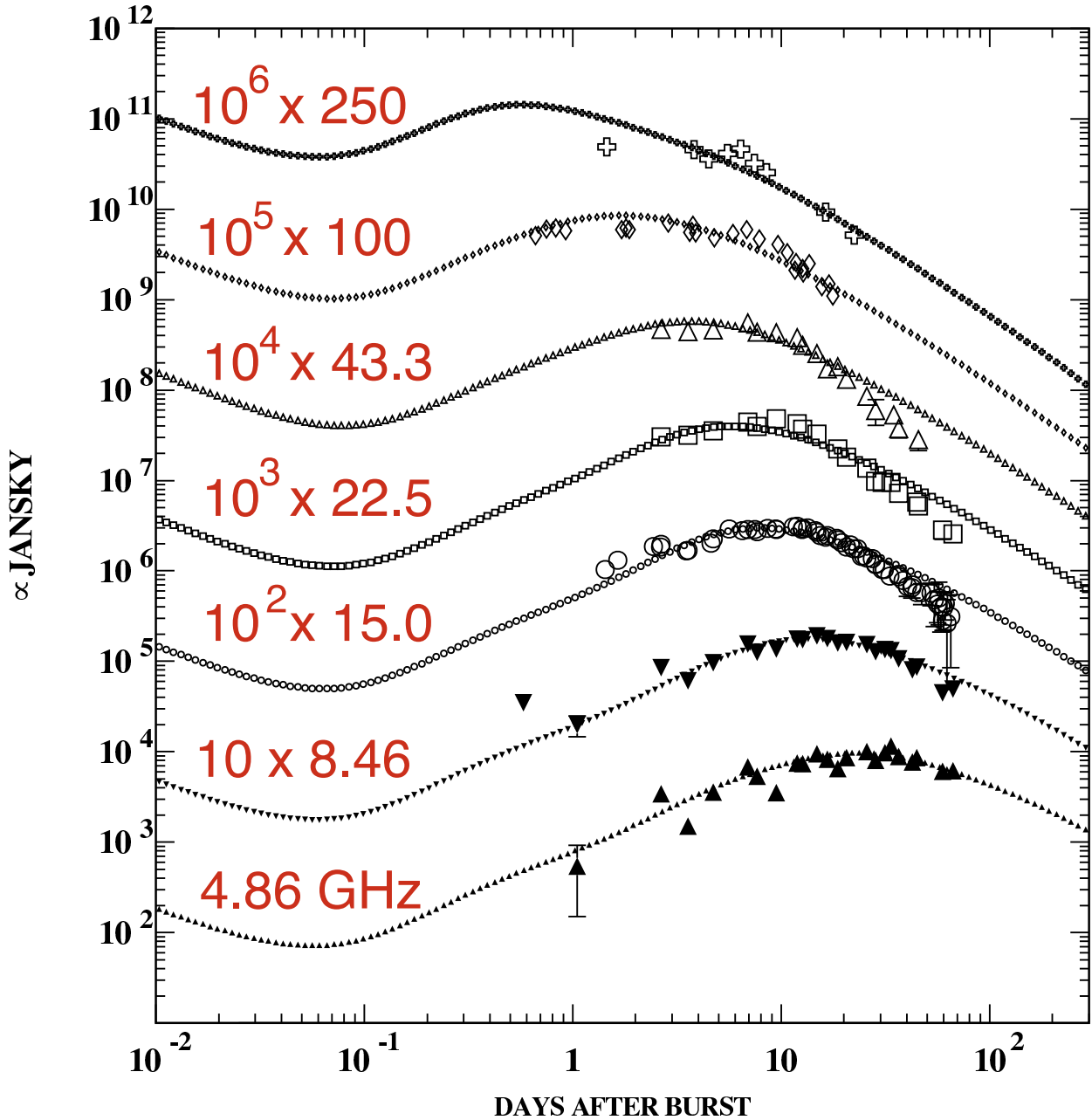}
\figcaption{The radio observations, ranging from 4.86 to 250 GHz,
of the AG of GRB 030329 (Sheth et al.~2003 and Berger et al.~2003)
and the CB-model ``second-round" broad-band fit with two CBs to these data
and the ones shown in Fig.~(\ref{figone}). 
The various frequencies have been scaled for
presentation. We attribute the ``residua" of this fit to having ignored
the ISM density inhomogeneities at $t>1$ day, as explained in the text.
\label{figthree}}
\end{figure}

\begin{figure}[]
\hskip 2truecm
\vskip -8cm
\vspace*{-0.4cm}
\plotone{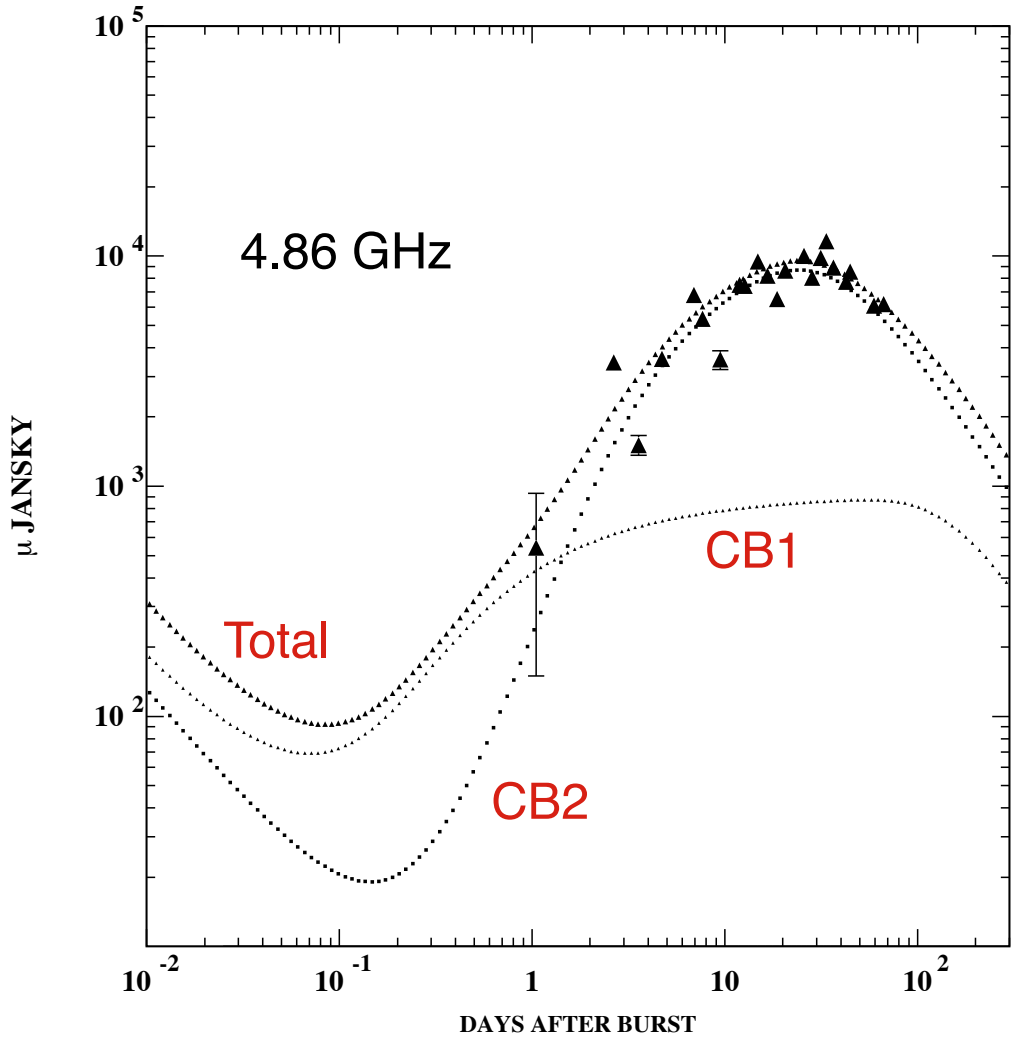}
\figcaption{Blow-up of the 4.86 GHz results
of Fig.~(\ref{figthree}). The best fit with two CBs is shown,
as well as their independent contributions. The cannonball
``CB2" dominates the radio AG at late times. 
\label{figfour}}
\end{figure}

The radio data at different frequencies and their comparison 
with the CB-model fit are shown in Fig.~(\ref{figthree}).
The fit to the  data at $\nu=4.86$ GHz is shown
Fig.~(\ref{figfour}), wherein the contributions of the
two CBs are separately shown. As 
for the optical data, Fig.~(\ref{figtwo}), we
conclude that the ``late" ($t>1$ d) radio data  
are also dominated by one of the CBs.

\section{The afterglow of GRB 030329: third round}

\begin{figure}[]
\hskip 2truecm
\plotone{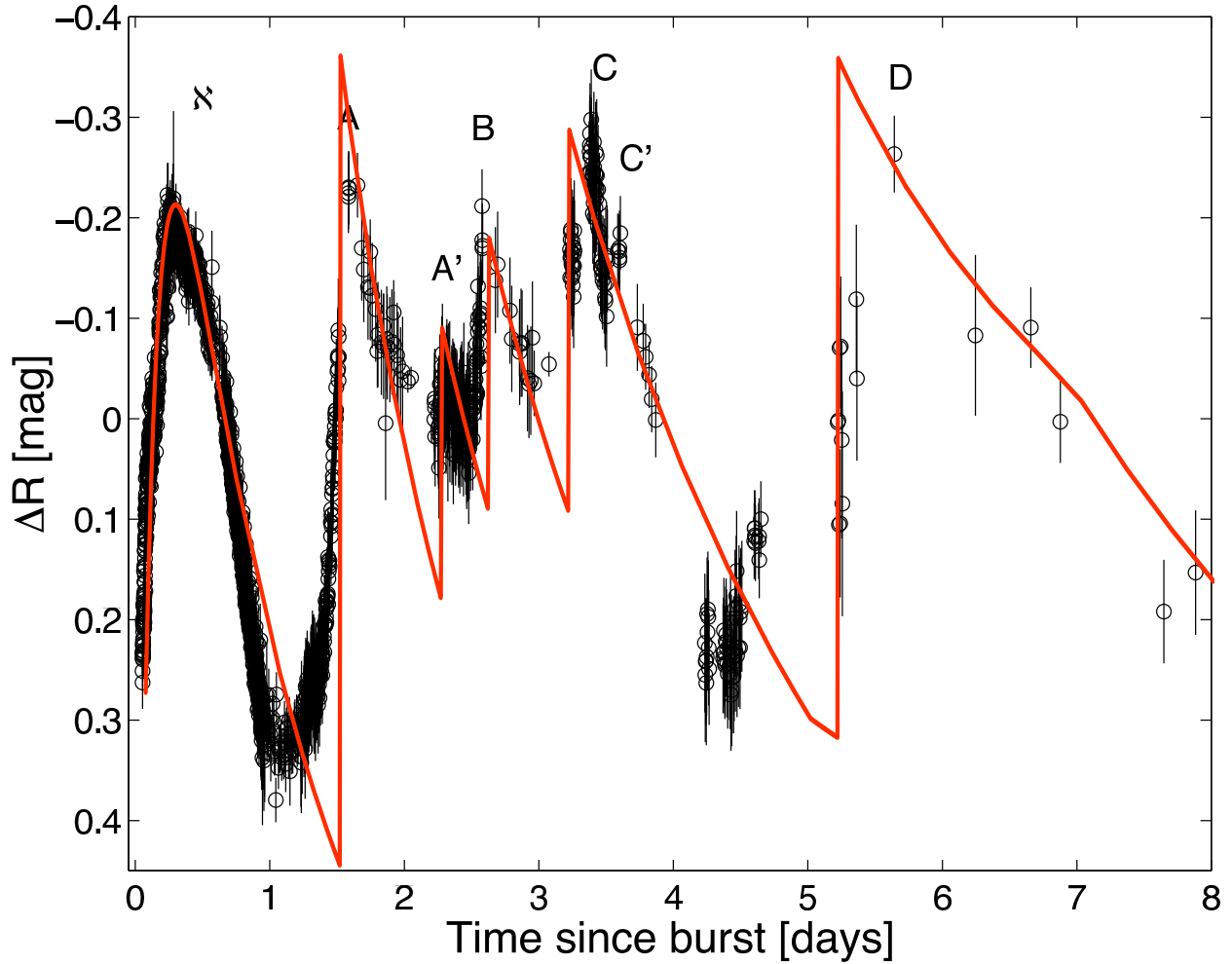}
\figcaption{The R-band AG of GRB 030329, shown in two
different ways. The ``residua" $\Delta R$ of the data (the black points and circles) 
are the observed magnitudes relative to a
broken power law of index $-\alpha$ jumping from $\sim 1.1$ to 
$\sim 2$ at $t\sim 5$ days (Lipkin et al.~2003).
The (red) line represents the residua, relative to the same broken power law,
of the ``third round'' CB-model fit described in the text.
In the CB model, the $\aleph$ feature is an artifact
of comparing a smoothly-varying curve (the data or the CB-model fit 
for $t<1$ day) to a power-law, which is theoretically unjustified at early times.
The other features are real and
have been modelled with the input density profile shown in Fig.~(\ref{figeight}).
\label{figseven}}
\end{figure}

\begin{figure}[]
\hskip 2truecm
\vskip -4.4cm
\plotone{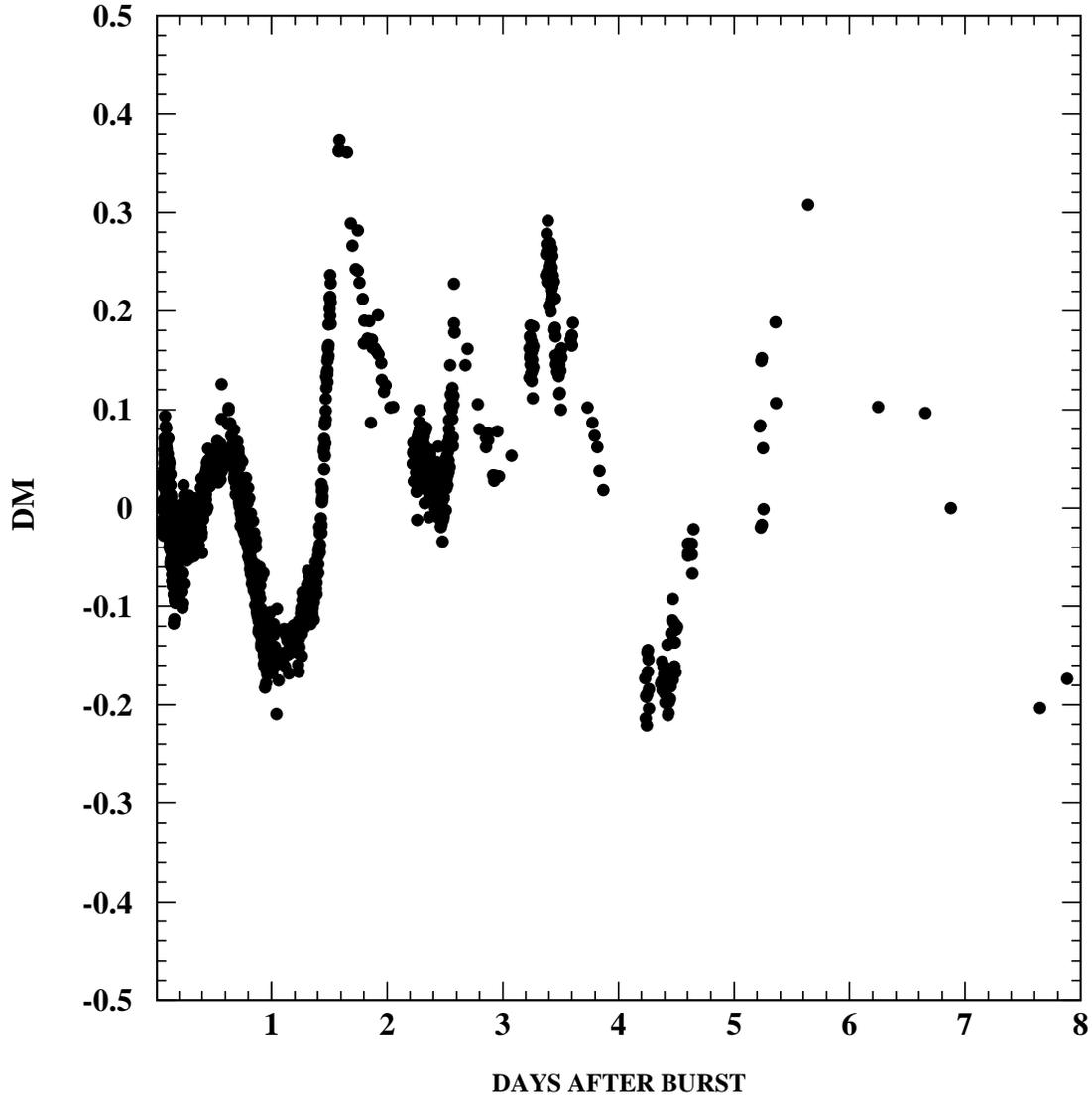}
\figcaption{The magnitude-difference ``residua" of the R-band
AG data of GRB 030329 (shown here 
without error bars) relative to the ``second-round" CB-model fit described
in the text, for which the ISM density is assumed to be a constant plus
a ``wind" contribution declining as $1/r^2$. The prominent $\aleph$ feature 
of Lipkin et al.~(2003) is absent, while the features at $t>1$ day are real.
\label{figten}}
\end{figure}

The rich structure of the AG of this GRB is shown in Fig.~(\ref{figseven}),
in which Lipkin et al.~(2003) have compared the R-band data to a ``reference" 
(a double power-law of indices $\sim -1$ and $\sim -2$, with the break at day 
$\sim 5$) by plotting the magnitude difference between the data and the reference.
The CB-model description of the optical and radio AGs in Figs.~(\ref{figone}) 
and (\ref{figthree}) satisfactorily reproduces the 
observed general trends. But the description is unsatisfactory in that
its residua are also significant, as we show in Fig.~(\ref{figten}). These
residua are not unlike those in Fig.~(\ref{figseven}), but for the 
absence of a prominent ``$\aleph$" feature, which we view as the result of the 
comparison of a
power law ---unmotivated at early times--- with the smoothly-varying data. 
The remaining 
features, particularly the more prominent ones occurring
after $t\sim 1$ day, require an explanation.

\begin{figure}[t]  
\vskip -.5cm
\begin{tabular}{cc}  
\epsfig{file=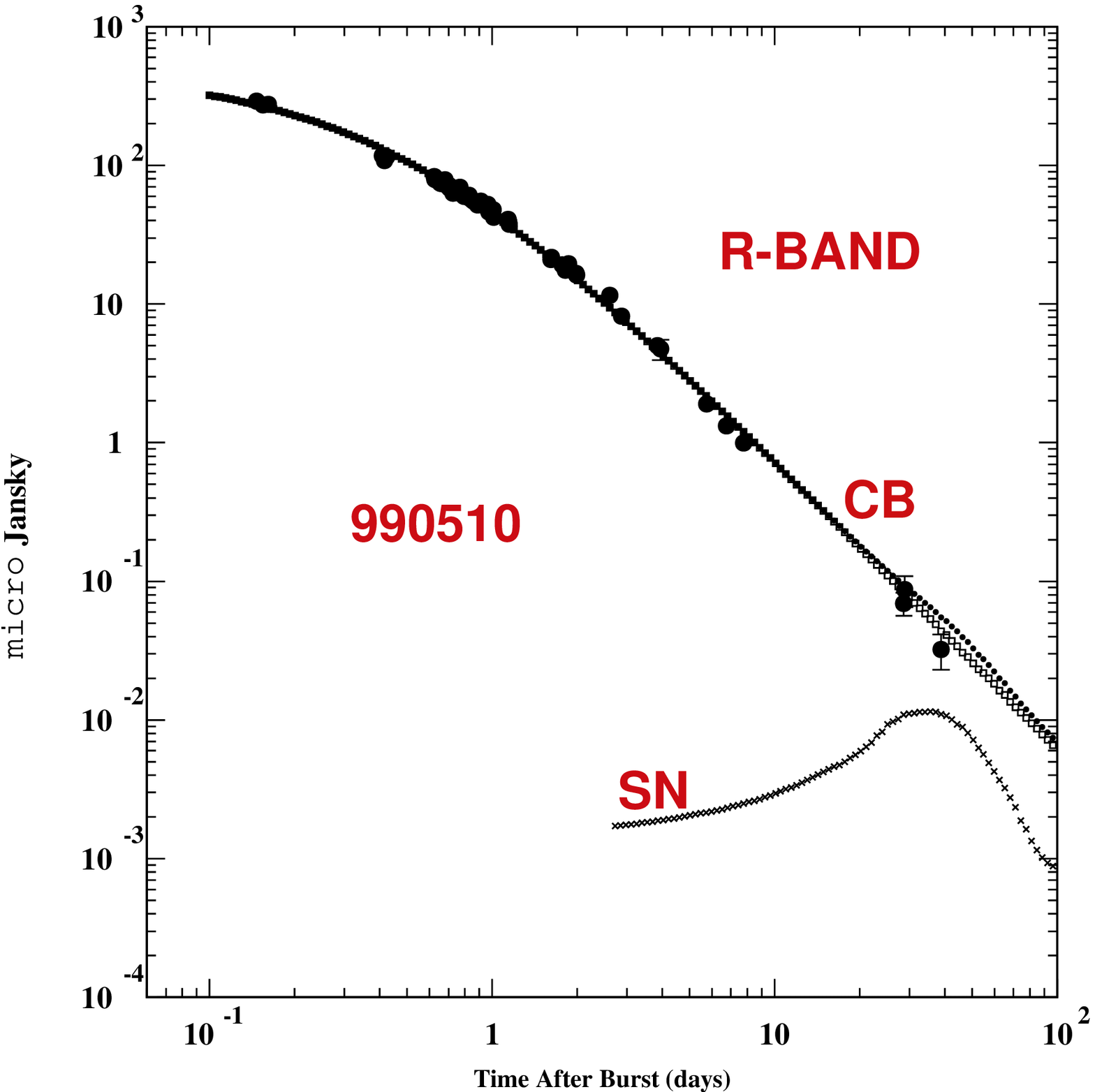, width=6cm}&
\vspace {.6cm}
\epsfig{file=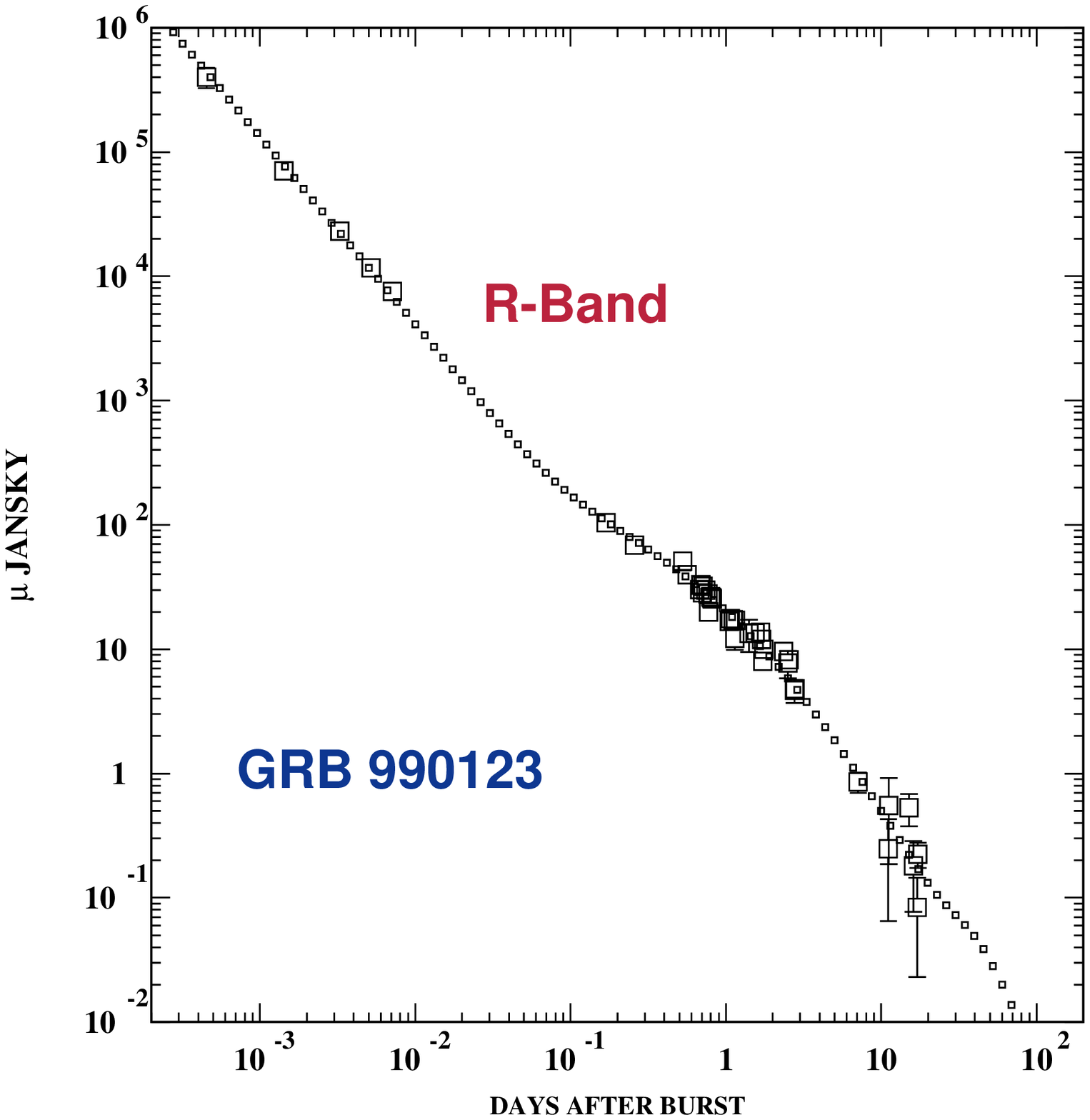, width=6.5cm}\\
\epsfig{file=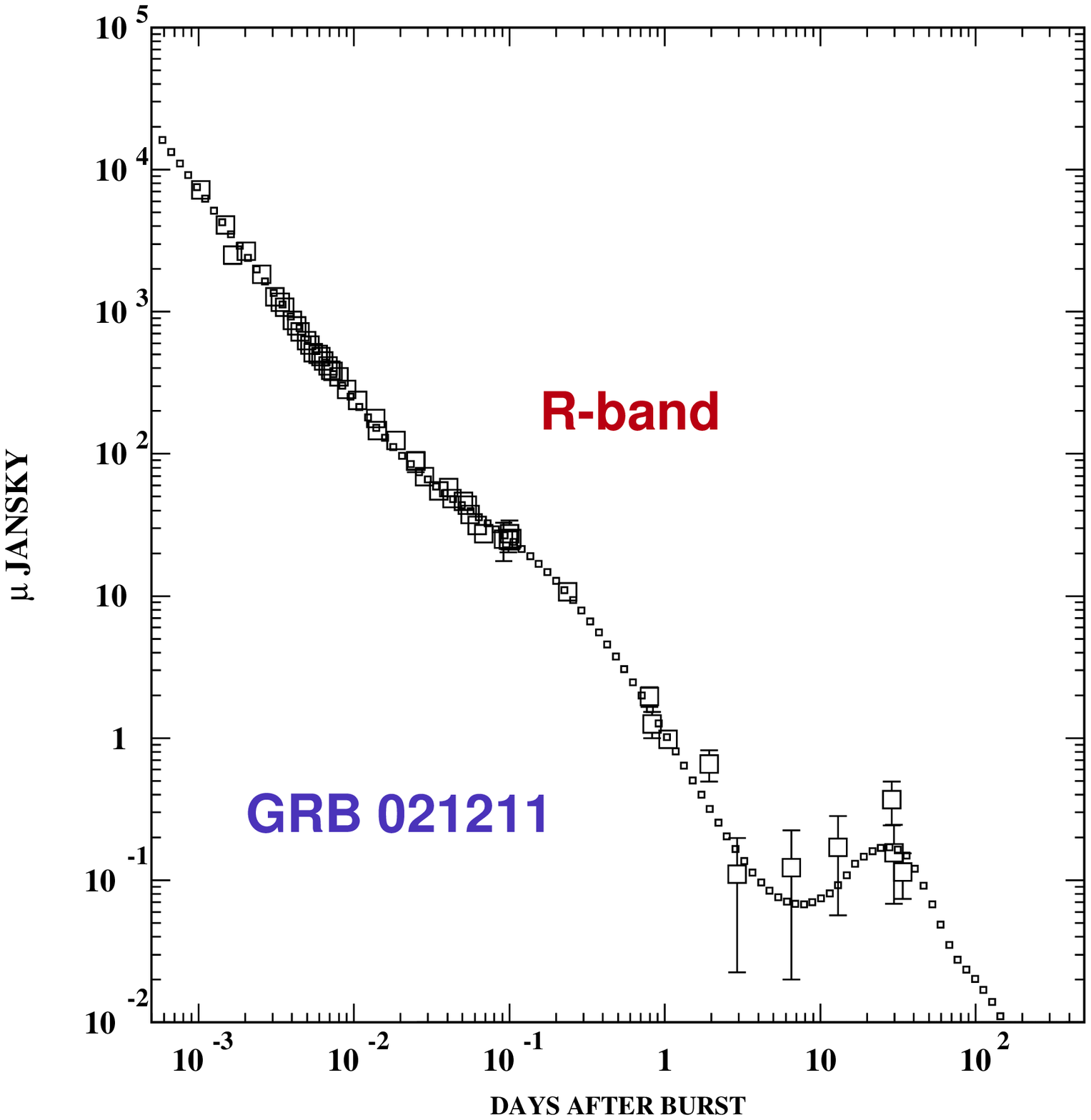, width=6.8cm}&
\vspace{.6cm}
\epsfig{file=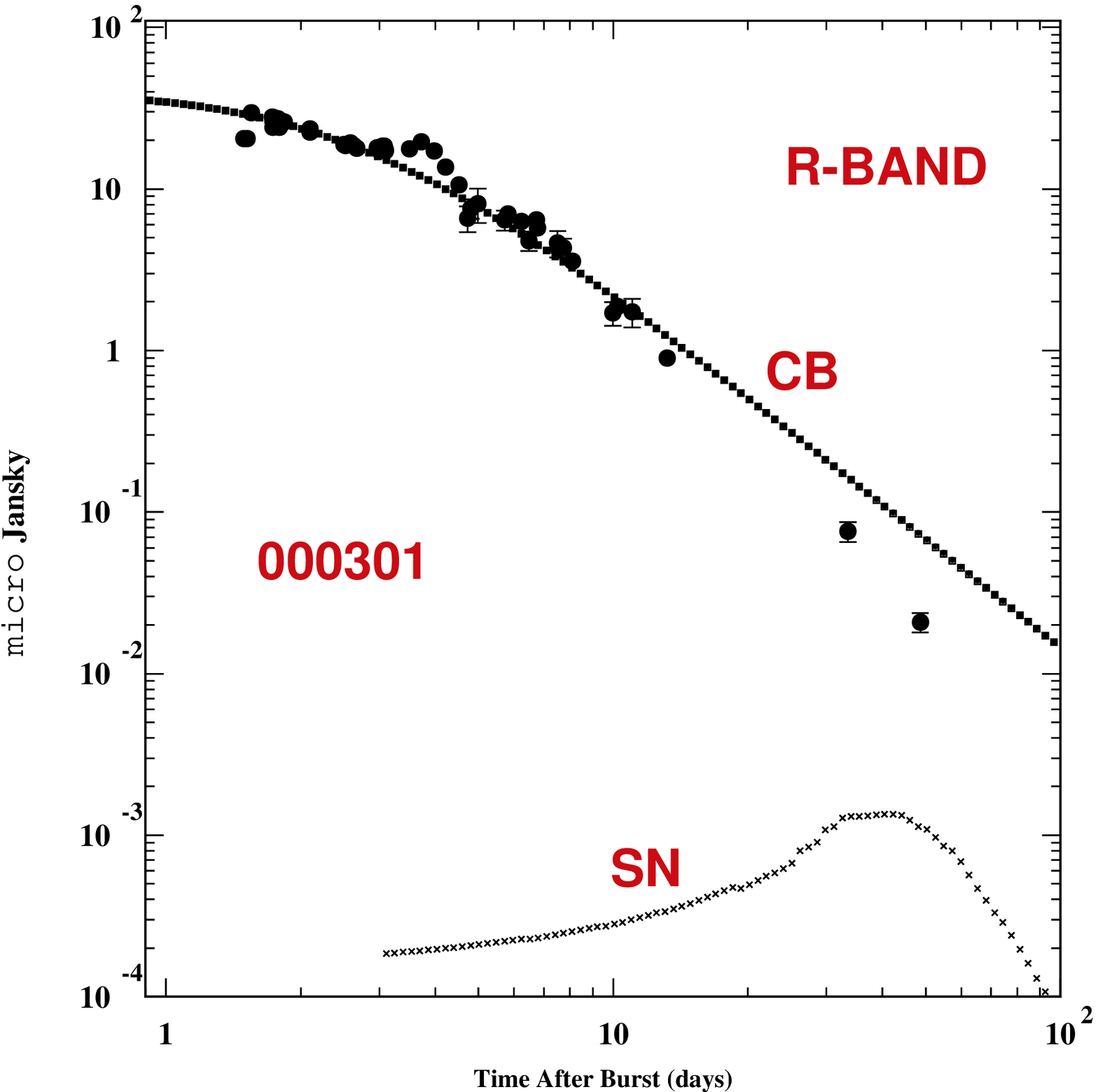, width=6cm}\\
\epsfig{file=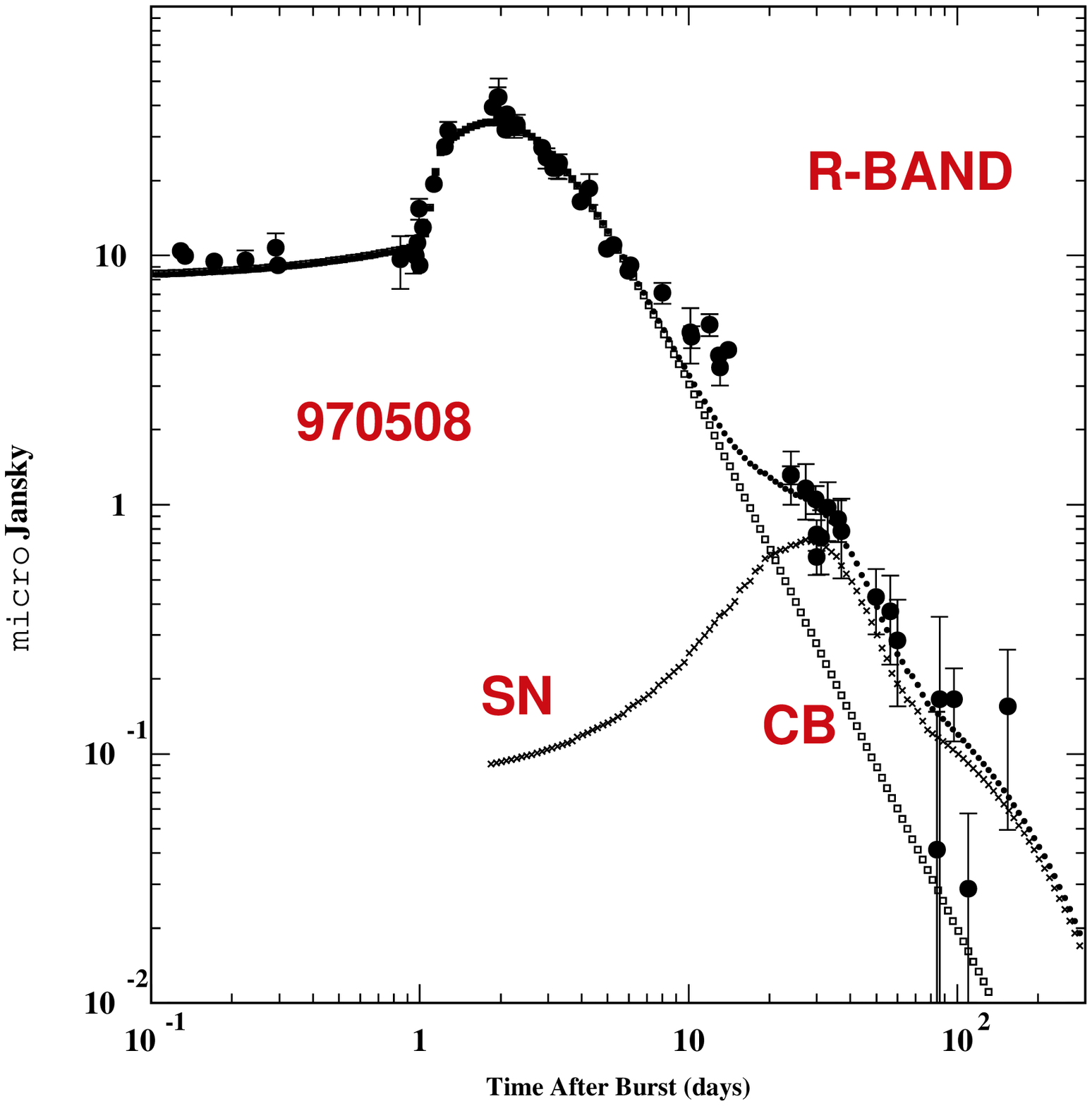, width=6.2cm}
\hspace{2cm}
&
\epsfig{file=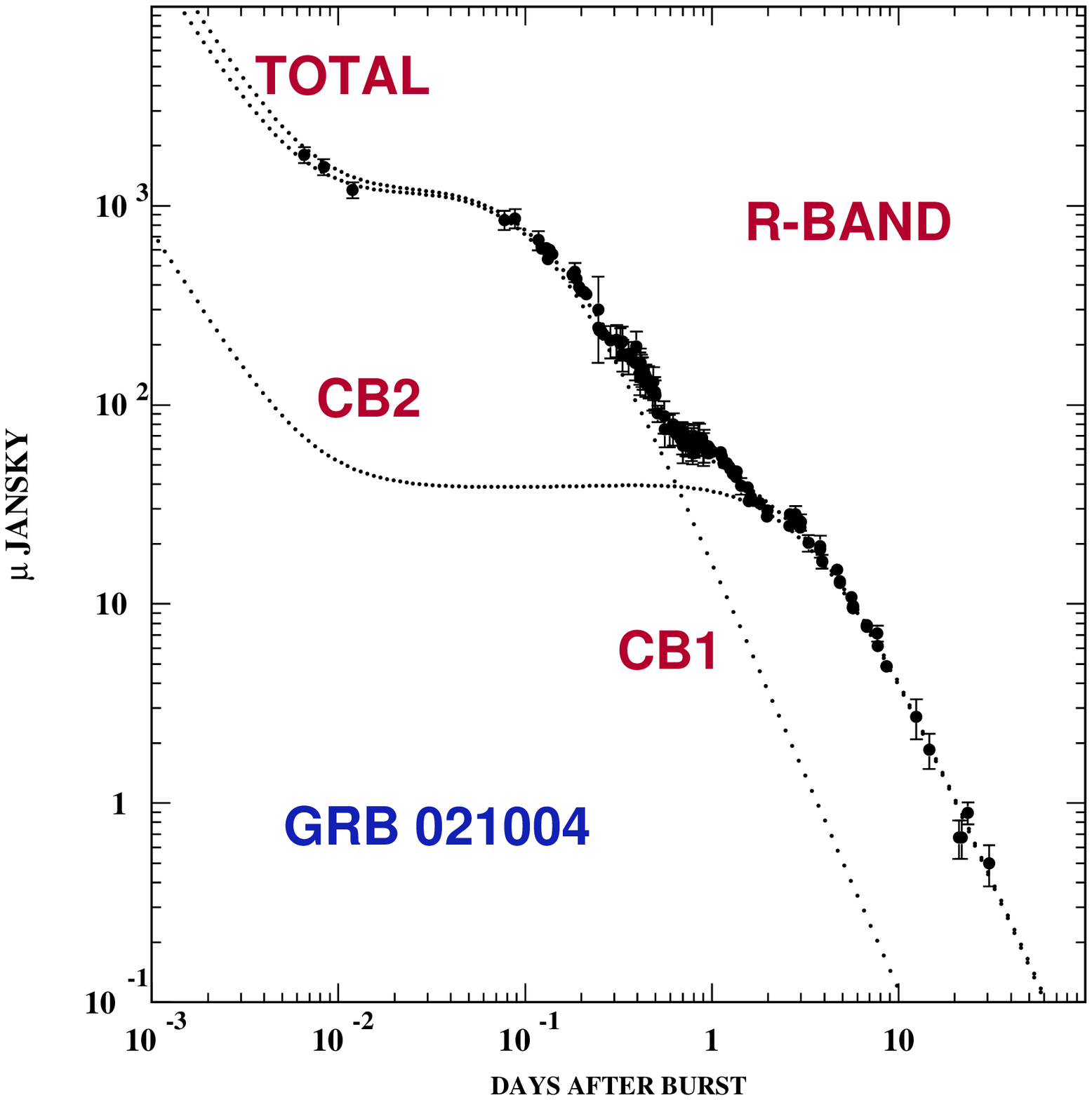, width=6.8cm}
\end{tabular}
\caption{R-band AGs of six representative
GRBs of interest to the analysis of GRB 030329,
and their CB-model fits. In a CB-model analysis, there is evidence
for a SN akin to SN1998bw (transported to the GRB location) in the AG of all 
GRBs in which such a contribution is discernible (in practice all GRBs 
with $z<1.1$).}
\label{tiriri}
\end{figure}

\begin{figure}[]
\hskip 1truecm
\vspace*{-0.4cm}
\plotone{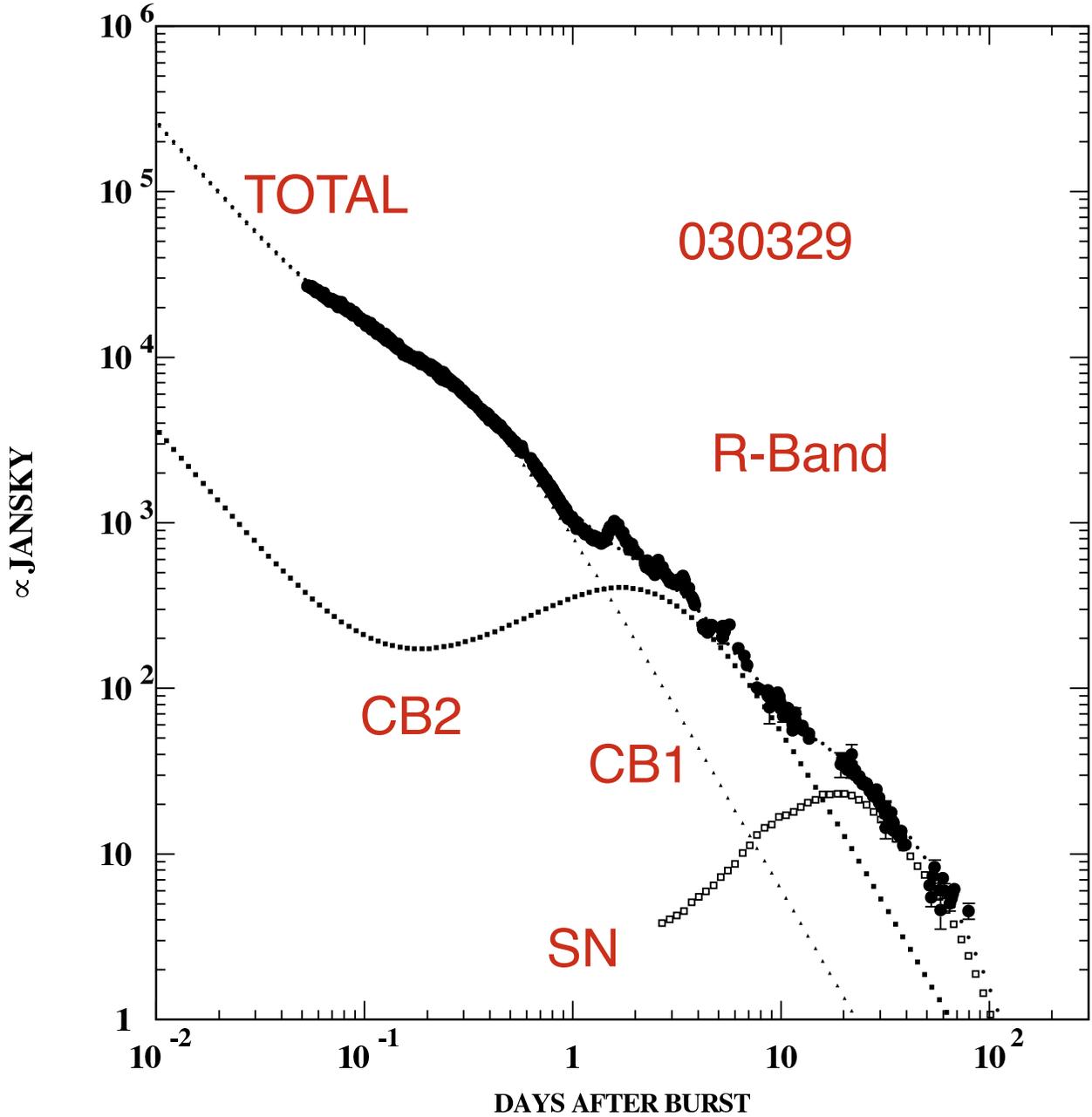}
\figcaption{Blow-up of the R-band results
of Fig.~(\ref{figone}).
The ISM density was
assumed to be a constant plus an additional ``wind'' contribution
decreasing as $1/r^2$. The wind contribution is
only significant at $ t\!<\!0.1$ days, after which the
CBs are more than 10 pc away from the progenitor.
This ``wind'' contribution is also seen in other AGs observed early enough,
e.g. GRBs 990123, 021211 and 021004, shown in Fig.~(\ref{tiriri}).
 The individual contributions of the two CBs
and of a SN akin to SN1998bw (at the GRB's redshift) are also shown;
the cannonball ``CB2" dominates the AG at late times.
We attribute the ``residua" of this ``second-round" fit to having ignored
the ISM density inhomogeneities at $t>1$ day, as explained in the text.
A fit with two distinct CB contributions was previously needed for
the description of GRB 021004, shown in Fig.~(\ref{tiriri}),
which also had a double-peak (two-CB) $\gamma$-ray light curve.
\label{figtwo}}
\end{figure}

The CB-model's fluence of Eq.~(\ref{afterglow}) is proportional to a 
power of the ISM's electron density $n_e=n_p$ and, in that sense, it
``traces" its local, instantaneous value. This tracing is not perfectly
{\it local}, because $F_{_{CB}}$ depends on $\gamma(t)$, and this
function reflects the integral effect of the ISM density that a CB
has swept through its prior voyage. To reproduce the features
still prominent in Fig.~(\ref{figten}), we must assume a given 
non-trivial density profile and solve Eqs.~(\ref{dgamma2}) to (\ref{tint})
explicitly to obtain the function $\gamma(t)$ to be used as
input in the expression for the fluence, Eq.~(\ref{afterglow}).
We have in the past done this for the AG of GRB 970508, 
as reproduced in Fig.~(\ref{tiriri}).

\begin{figure}[]
\hskip 2truecm
\plotone{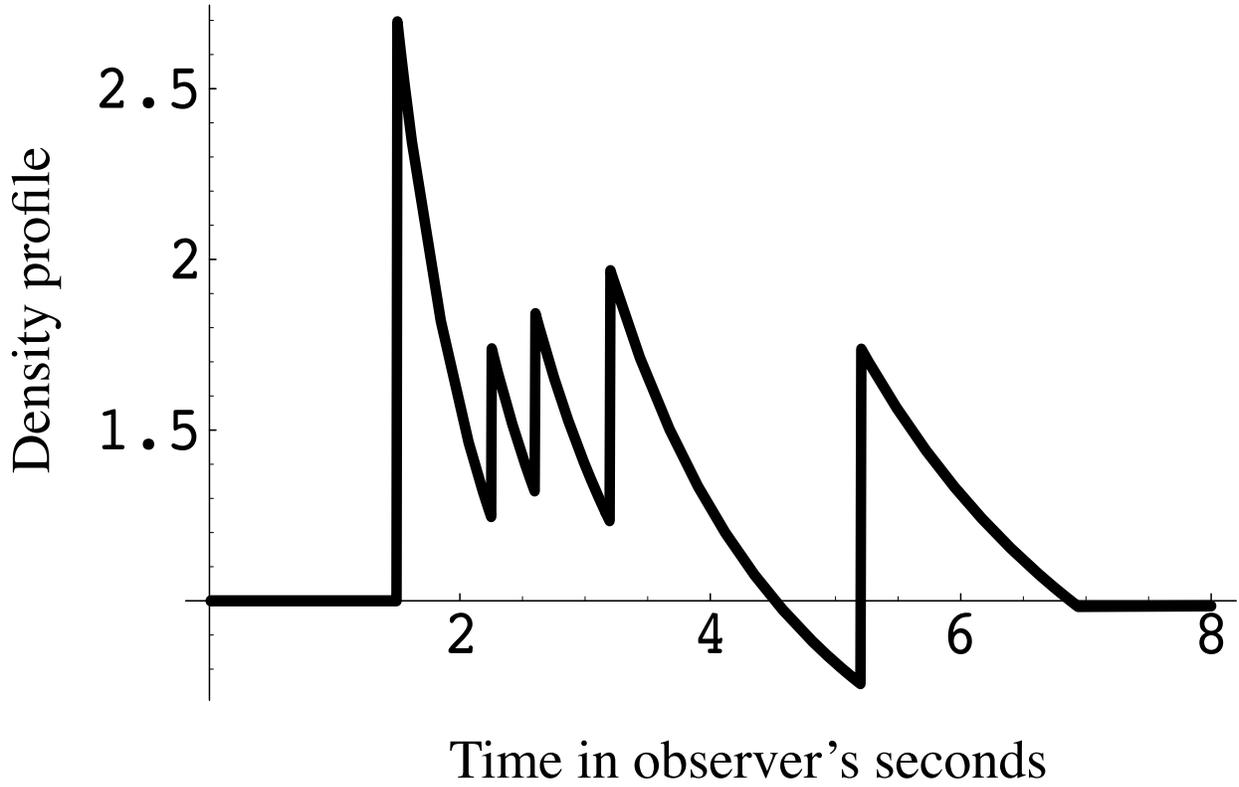}
\figcaption{The ``structured" density profile: the
ISM density variations assumed in the ``third round" CB-model
fit to the R-band AG of GRB 030329
(relative to the smooth ISM density of the ``second round" fit, assumed to be
a constant plus a ``wind" contribution decreasing as $ 1/r^2$).
\label{figeight}}
\end{figure}

The CBs of GRBs 970508 and 030329, as we shall see explicitly for the 
second one, are at a distance $x\sim 100$ pc from the SNe that emitted 
them, at an observer's time $t\sim 1$ day.
At a distance of that order, we expect the CBs to be exiting the superbubble
where the SN explosion is very likely to have taken place.
As they exit it, they may encounter successive density 
inhomogeneities produced by a succession of past SN explosions
and stellar mass ejections,
which created the bubble in the first place. We have considered
the {\it structured density profile} shown (as a function of observer's time) in 
Fig.~(\ref{figeight}), in which successive ``onion peals" appear
as abrupt density increases, followed by a decline proportional
to $1/r^2$ (or, approximately, $1/t^2$), analogous to the
closer-by profile of a ``wind-fed" circumburst material. With such a 
profile of overdensities we obtain a description
of the R-band AG whose residua (relative to the reference broken 
power-law introduced by the observers) are shown in Fig.~(\ref{figseven})
as the (red) line. The parameters of this fit are only slightly different
from the ones for a constant $n_e\approx n_p$, but for a somewhat larger 
initial value of $x_\infty$ for the CB dominating the AG at late times.
The $\aleph$ feature is fairly well reproduced with the 
density profile of Fig.~(\ref{figeight}), which is constant up to day 
$\sim 1.5$. The feature is, in this sense, a fake. The description of
the later-time features, however, requires the structured density
profile of Fig.~(\ref{figeight}).

\begin{figure}[]
\vskip -6truecm
\epsfig{file=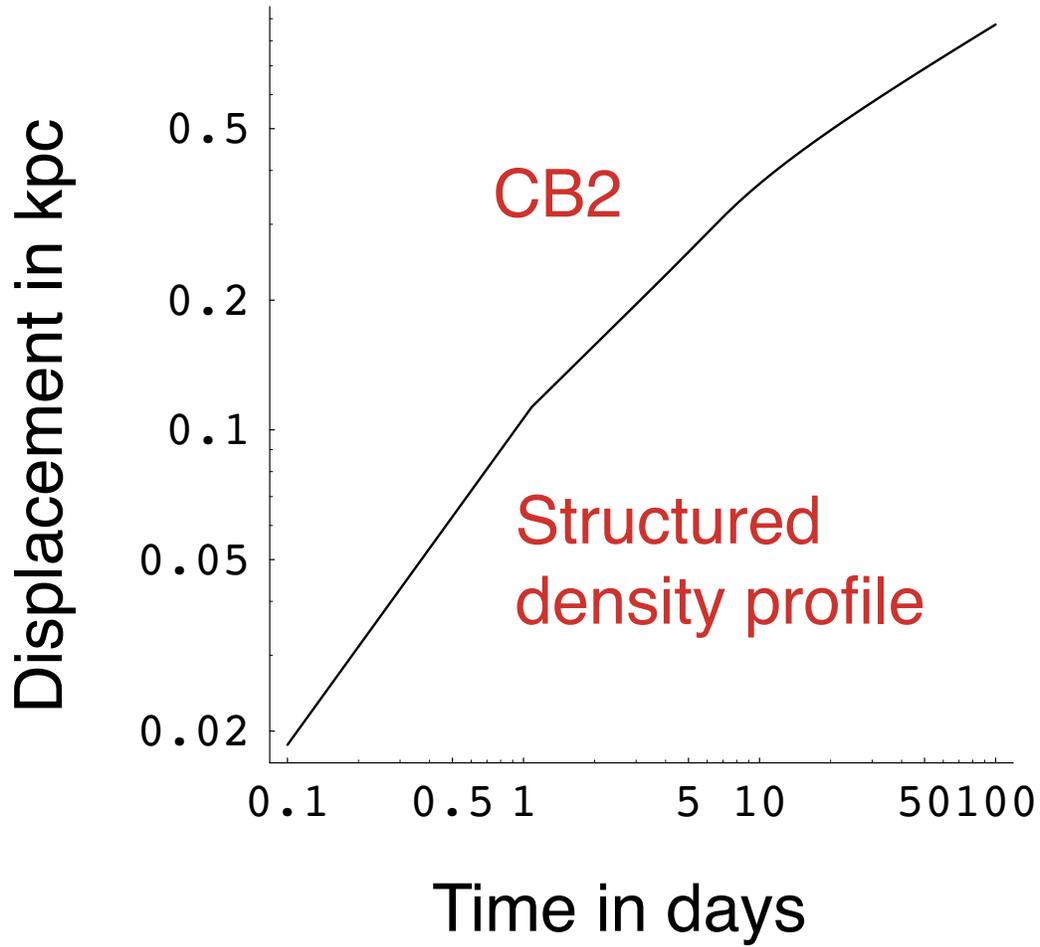, width=14cm}
\figcaption{The distance in kpc, as a function of observer's time,
of the CB which dominates the AG of GRB 030329 at $t>1.5$ days, and
whose displacement in the sky is largest. The prediction is for
the ``third round" fit described in the text, with the structured
density profile of Fig.~(\ref{figeight}), required to explain the
features of the AG.
\label{figkpc}}
\end{figure}

Clearly the many-parameter exercise we just described is not a ``fit": 
it could easily be improved to obtain an even better description,
and it is certainly not as ``predictive" as our prior fits to the early
part of the AGs of GRBs 990123 and 021211,  from which we claimed to
have successfully traced the magnitude and $1/r^2$ profile of the 
close-by circumburst density (Dado et al.~2003d). Yet, the description is 
phenomenologically
satisfactory, in the sense that we can once again claim that, in the
CB model, the shapes of AGs provide interesting and consistent
{\it quasi-local} tracers of the ISM density through which the CBs travel.
The actual distance $x(t)$ from the SN at which the late-AG-dominating CB in 
GRB 030329 was, as seen at a given observer's time $t$, is shown in
Fig.~(\ref{figkpc}), constructed with use of Eq.~(\ref{tint}). Indeed,
at the time at which the density inhomogeneities are apparent
($t$ between $\sim 1$ and $\sim 5$ days) the CB is 100 to 200 pc
away from the SN, a reasonable radius for the overdensities 
surrounding a superbubble.

Given the need to introduce density variations to describe the detailed
behaviour of the AG, the question arises whether or not two CBs
---as opposed to just one--- are needed. Our attempts to describe the
AG with just one CB and a variety of density profiles failed. The reason
is simple and unavoidable: the density increase required to produce
the AG's ``shoulder" at $t>1$ day inescapably increases the rate
at which $\gamma(t)$ decreases with time. This entails a fast
fall-down of the fluence, following soon after the initial increase
produced by the local density increase. The resulting light-curves
rise and fall fast ---like the one shown in Fig.~(\ref{tiriri})
does in the case of GRB 970508--- and they fail in the description of
the AG of GRB 030329. Thus, two CBs appear to be necessary
to understand the AG of GRB 030329,
as they are for the description of its two-pulse GRB structure, 
shown in Fig.~(\ref{GRB}).

\section{Density variations, as observed at different frequencies}

We are primarily concerned in this note with the putative superluminal
motion of the CBs of GRB 030329, and not with a complete description
of its wide-band AG. Thus, we have limited our detailed analysis with
a structured density profile to the R-band AG, as in Figs.~(\ref{figseven}),
the complete broad-bend analysis with density variations
being very laborious.

The outcome of the discussion that follows is that the structure seen
in the R-band AG should be {\it almost} achromatic for data ranging
from X-ray to NIR frequencies, while the structures should progressively
disappear at lower and lower radio frequencies.  The X-ray and optical
AGs are not perfectly achromatic because the injection-bend frequency 
of Eq.~(\ref{nubend}) ``crosses'' a given frequency at different times.
But the crossing of the optical frequencies typically occurs at $t\sim 1$
day and the light curves become increasingly achromatic thereafter.
The radio AGs, on the other hand, are predicted to be strongly chromatic
(Dado et al.~2003a). To be more specific, we quote from the
mentioned reference:

{\it ``Electrons that enter a CB with an injection Lorentz factor  $\gamma(t)$
are rapidly Fermi accelerated... On a longer time scale, they
lose energy by synchrotron radiation, and their
energy distribution evolves... 
Electrons with a large $\gamma\sim {\cal{O}}\,[\gamma(t)]$ emit
synchrotron radiation, with no significant time-delay, 
at the observer's optical and X-ray wavelengths.
But the emission of radio is delayed by the time it takes the electrons
to ``descend'' to an energy at which their characteristic emission
is in the observer's radio band... 
Thus, the optical and X-ray AG emission
starts ... a few observer's seconds after the corresponding $\gamma$-ray
pulse. The radio signal, on the other hand, must await 
a time $ \Delta t$}  {\it for the cumulated electrons to cool down."}
$\Delta t$ was explicitly estimated in Dado et al.~(2003a) to be
of ${\cal{O}}(1)$ day.

Applied to the case at hand of a varying ISM density, what this means
is that the features observed in the optical AGs should be smoothed
in the radio over intervals of ${\cal{O}}(1)$ day. These intervals are
frequency-dependent: longer for the lower radio frequencies.
The expected trend of features that progressively disappear as
the frequency diminishes is precisely the
trend observed in the radio data reproduced
in Fig.~(\ref{figthree}). The ``second-round'' fit shown in the figure
systematically overestimates the late results, also a consequence of
having ignored the density enhancements.

\section{The superluminal motion in GRB 030329}

The location of the source of the AG of this GRB has been
followed in detail by VLBA observers from day $\sim$ 3 
to day $\sim$ 84 after burst (Taylor et al., to be published).
The associated SN2003dh dominates the optical AG
after day $\sim 10$, but is negligible in the radio AG at
all measured times. Thus, the location of the radio source(s)
is the location of the CB(s), expected to be extremely superluminal
in the CB model.  

  The values of the fit observer's angles are very similar for the two CBs: 
$\theta[1]=2.2$ mrad, $\theta[2]=2.3$ mrad. The values
of the initial Lorentz factors are not so different: $\gamma_0[1]=1037$,
$\gamma_0[2]=1606$. But the values of the deceleration parameter
are very different: $x_\infty[1]=0.033$ Mpc, $x_\infty[2]=0.37$ Mpc
(these numbers refer to the initial $x_\infty$ as in Eq.~(\ref{xinf}), 
still unaffected by the density variations occurring at $t>1$ day).
This is the main reason why the contribution of CB1 to the AG 
decays much faster 
with time than that of CB2, as implied by Eq.~(\ref{cubic0})
and illustrated in Figs.~(\ref{figtwo},\ref{figfour}). 
It is also the reason why the superluminal motion of CB1 is
much slower than that of CB2, as implied by Eq.~(\ref{Deltaalpha}).
We are therefore interested in the fastest-moving 
and late-AG-dominating cannonball: CB2.

\begin{figure}[t]  
\vskip -8cm
\begin{tabular}{cc}  
\epsfig{file=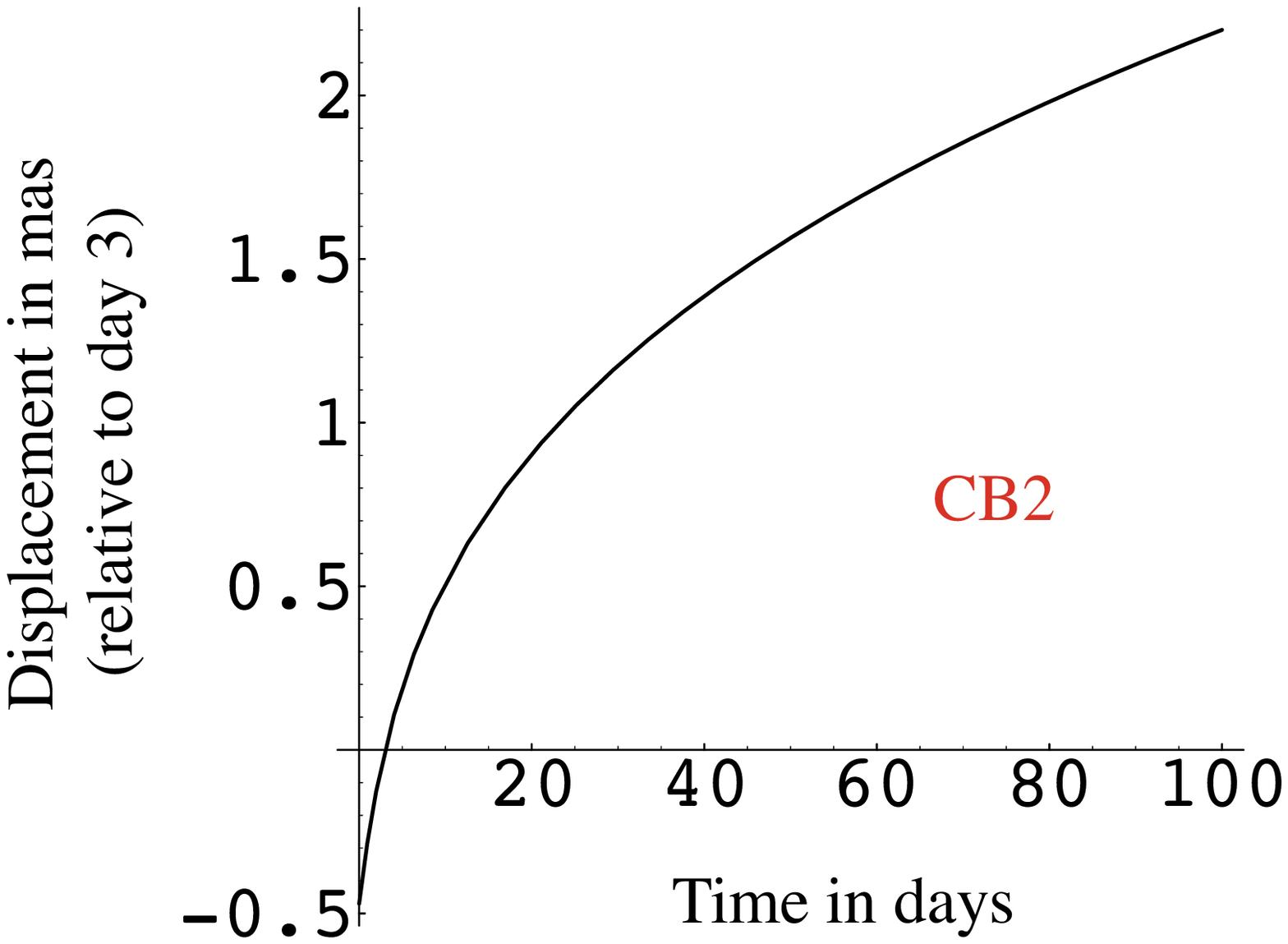, width=13cm}\\
\epsfig{file=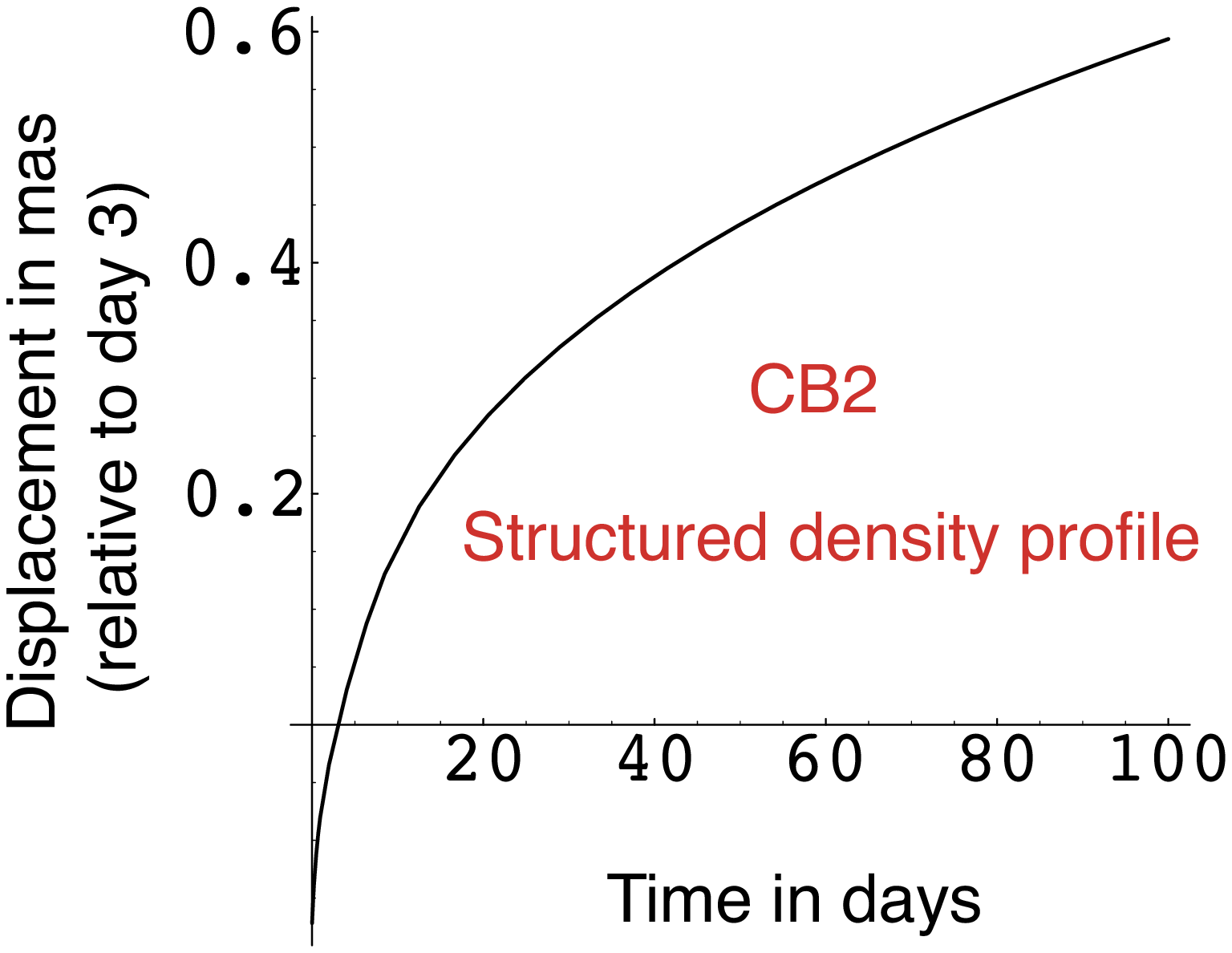, width=10.3cm}
\end{tabular}
\caption{The predicted angular displacement 
in the sky (in mas) as a function of observer's time, of the CB which 
dominates the AG of GRB 030329 at $t>1.5$ days, and
whose displacement is largest. The displacement is shown relative to
the CB's position at day 3. The upper panel is the prediction of
Dar \& De R\'ujula (2003b), wherein the effect of the observed ``features" of 
the AG was ignored. The lower panel is the prediction
for the ``third round" CB-model fit described in the text, in which the
cited effect is taken into account via the structured density profile of 
Fig.~(\ref{figeight}).}
\label{tororo}
\end{figure}

In the absence of the density fluctuations of Fig.~(\ref{figeight}), the
predicted angular displacement $\Delta\alpha(t)$ of CB2, as given
by the integrated form of Eq.~(\ref{Deltaalpha}), would be
that of the upper panel of 
  Fig.~(\ref{tororo}). It is on the basis of our ``first-round"
constant-density fit that we predicted such a displacement in 
Dar \& De R\'ujula (2003b). In the presence of the observed density fluctuations,
the predicted $\Delta\alpha(t)$ is shown in 
the lower panel of Fig.~(\ref{tororo}),
the result of the integration of the 
velocity $V_{_{CB}}$ in Eqs.~(\ref{supervelocity},\ref{Deltaalpha}), 
obtained from the full-fledged
determination of $\gamma(t)$ from Eqs.~(\ref{dgamma1}) to 
(\ref{tint}). The density fluctuations are dominantly density
enhancements, implying a faster deceleration of the CB. 
Thus the significantly reduction in the predicted apparent
superluminal motion. Alas, this prediction of the CB-model is,
in the case of GRB 030329, harder to test than we originally
thought.

\section{Caveats?}

In its current form, the theory of AGs in the CB model is based on an
analogy with high-resolution observations of relativistic jets emitted by
systems such as the microquasars SS 433 
(http://chandra.harvard.edu/ photo/2002/ 0214/ index.html;
http:// chandra. harvard. edu/ press/04$_{-}$releases/ press$_{-}$010504.html),
GRS 1915+105 (Mirabel \&
Rodriguez 1999; Dhawan et al.~2000) and XTE J1550-564 (Corbel et al.~2002);
and active galactic nuclei, e.g.~the quasar 
Pictor A, whose emitted CBs, as seen at X-ray wavelengths,
seem to stop expanding laterally shortly after ejection, and to travel for
hundreds of kpc, before they finally stop and blow up (Wilson, Young \&
Shopbell 2001; Grandi et al.~2003). The CB-model's simple explanation for
this surprising fact is the following (Dado et al.~2002a).  The ambient
protons intercepted by a CB in its voyage encounter its inner chaotic
magnetic field and are reemitted roughly isotropically (in the CB's rest
system). This implies an inwards pressure that stops the CB's expansion
and that is asymptotically equal to the pressure of the CB's inner
magnetic field, which is thereby calculable. The corresponding
deceleration law (at constant radius $R_\infty$ and ISM density $n_p$) 
is that of Eq.~(\ref{cubic0}).

The magnitude and time-dependence of the magnetic field determined as in
the previous paragraph plays a crucial role in the prediction of the
broad-band AG spectra, which in the CB model is extremely simple,
parameter-thrifty, and successful (Dado et al.~2003a, 2003e), particularly
in comparison with its standard counterpart (see, e.g.~Granot \& Sari
2002).  In spite of their phenomenological success, our assumptions
leading to calculable CB radii and magnetic fields are very bold, and are
no doubt oversimplifications. To investigate their solidity, in Dado et
al.~(2002a) we also studied a different ansatz: that the CB's radius $R$
would keep on increasing constantly with $dx$, with the ISM density still
kept approximately constant. This ansatz, which corresponds to a much
faster CB deceleration than that of Eq.~(\ref{cubic0}), failed miserably.
But in between the two extremes (a non-increasing and a steadily
increasing CB's radius), it is clear that there should be choices of
$R(x)$ and $n_p(x)$ that are phenomenologically satisfactory and yet,
correspond to a {\it faster} CB deceleration than the simple one dictated
by Eq.~(\ref{cubic0}) at constant $R\times n_p$. For such choices the
simplest predictions of superluminal velocities ---that we have
discussed--- would result in overestimates. But it would
be premature to modify the simple predictions
of the CB model, which ---so far--- are succesful.

\section{Discussion}

In the internal/external shock model of GRBs, the variability of AG
light curves ---the set of ``features" that we have discussed--- is 
attributed to
patchy shells and refreshed shocks (see, e.g.~Piran, Nakar \& Granot
2003; Granot, Nakar \& Piran 2003  and references therein). In the case of 
GRB 030329, the features
are attributed to delayed collisions between late, relatively low-$\gamma$
shells and the earlier higher-$\gamma$ shells having decelerated in
their interactions with the ISM. The timing and the Lorentz factors are
chosen so that the early fast shells collide in a matter of (observer) seconds
to produce the GRB peaks, while the slow shells catch up with the
decelerated earlier ones at times of order days: some five order of
magnitudes longer. The required amount of fine-tuning is considerable.
The late shells bring in an energy injection an order of magnitude larger
than that of the original blast wave. It would be interesting to know whether
their collisions with the slowed-down shells are expected to give rise
to GRB pulses as well, in which case GRBs should be ``repeaters".

In the CB model, as we have seen, all that is required to explain the AG
``features" are ISM density inhomogeneities occurring at a natural
distance from the parent SN: that corresponding to the radius of
a superbubble. No new phenomena must be invoked and no GRB
parameters must be fine-tuned with special care. 

\section{Conclusions}

The data on GRB 030329 are now sufficiently complete to
allow for a detailed prediction of the motion of the source of its AG
---allegedly superluminal in the CB model.
The $\gamma$-ray light-curve and the
optical AG require the presence of two CBs, one of
which dominates the AG at late times. The parameters
needed to predict the motion of the two CBs in the sky
are determined by the optical data, so that the individual
motion of each CB is predicted. The main result of this
paper is the prediction of the sky-motion of the fastest-moving
CB, which dominates the late AG, and is shown in 
the lower panel of Fig.~(\ref{tororo}).

 The main caveat concerning a putative superluminal signature 
 (Dar \& De R\'ujula 2000, 2003b) concerns the input to the estimate
 of its magnitude. Indeed, we have shown in this paper that the
 predictions are very sensitive to the details of the density profile
 of the ISM. But we are encouraged by the fact that in the
 only case in which the superluminal jets of CBs made by a 
 core-collapse SN could be seen, they were seen. Indeed, observations
 of SN 1987A (Ninenson and Papaliolios, 1999) showed two sources,
 moving in opposite directions along the SN's axis at (real)
 velocities compatible with the speed of light and at an apparent
 superluminal velocity for the approaching source. Mercifully,
 the jet of that SN was not pointing in our direction (Dar, Laor \& Shaviv
 1998; Dar \& De R\'ujula 2001b).

Regarding the search for a superluminal
motion, we learned by reading the e-version of NYT 030529
(the New York Times of that date, in GRB's parlance)
 that, according to Dale Frail {\it ``[Our observations] are sufficient to 
rule out predictions of the cannonball model"}. We have shown that,
indeed, the observations of complicated features in the optical
AG of GRB 030329 imply that our earlier results (Dar \& De R\'ujula 2003b)
---which ignored the presence of these features--- constituted an
overestimate of the predicted superluminal displacement. In this sense, 
Frail was right in stating that the observations ruled out the 
{\it predictions}, as opposed to the model itself.

In a setting more scientific than the NYT, 
Bloom et al.~(2003) state: {\it ``Owing to the proximity
and bright radio emission, high-resolution ($\sim1$ pc) {\it Very Long
Baseline Array} imaging of the compact afterglow was used by
Frail (2003) to} {\bf unequivocally disprove} {\it the cannonball model
 for the origin of GRBs."} The emphasis is ours. We have seen that
these news of the death of the CB model may have been premature. 
Even though Mark Twain eventually died for sure, the CB model 
---though probably not immortal--- is still in an excellent shape.
Yet, trying to disprove the best available model(s), or even the proof
of a difficult theorem, is the acceptable standard attitude in many a realm of the
exact sciences. In this sense, the apparently strong
motivation of the quoted observers to disprove the CB model is
---in our opinion, and regarding this particular model---
the healthiest of all possible attitudes.

\noindent
{\bf Acknowledgements.} One of us (A.~De R.) is indebted to the
Physics Department and Space Research Institute of Technion
for its hospitality. This research was supported in part by
the Helen Asher Space Research Fund for research at the Technion.

\section{Appendix: The CB model}

In this model, the {\it engines} generating long-duration GRBs
and XRFs are ordinary core-collapse SNe.
Following the collapse of the stellar core into a neutron
star or a black hole, and given the characteristically large
specific angular momentum of stars, an
accretion disk or torus is hypothesized to be produced around
the newly-formed compact object, either by stellar material originally   
close to the surface of the imploding core and left behind by the
explosion-generating outgoing shock, or by more distant stellar matter
falling back after its passage (De R\'ujula 1987). A CB is emitted, as
observed in microquasars, when part of the accretion disk
falls abruptly onto the compact object  (e.g.~Mirabel \& Rodrigez 1999;
Rodriguez \& Mirabel 1999 and references therein). The 
CBs of a GRB are assumed to me made of ordinary hydrogenic 
matter. This is also in analogy to microquasar CBs, observed to be made of 
ordinary matter, as opposed
to the standard-model contrived mixture of baryons and $e^+e^-$ pairs.
Indeed, blueshifted and redshifted H, He, metal and heavy-element 
optical and UV lines were detected from the approaching and receding 
CBs emitted  by the microquasar SS433; see, e.g.~Eikenberry et al.~(2001),
Gies et al.~(2002).

In the CB model, SN1998bw, associated with GRB 980425, is an ordinary
core-collapse SN (Dar and Plaga 1999; Dar and De R\'ujula 2000): its
``peculiar'' X-ray and radio emissions were not emitted by the SN, but
were part of the GRB's AG (Dado et al.~2002a, 2003a). The GRB appeared to
be underluminous because it was viewed further off axis than
others\footnote{This conclusion has been recently ``standardized''
(i.e.~incorporated into standard GRB models without proper reference)
by Granot, Panaitescu, Kumar \& Woosley~(2002).}
(e.g., Dar and De R\'ujula 2000).
Thus, it makes sense in this model to expect an association between
(long) GRBs and SN1998bw-like engines, transported to the GRB
location (Dar 1999). The a-posteriori analysis of Dado et al.~(2002a)
of all the available data supported the conclusion
that {\it all} (long) GRBs are indeed associated with such 
SNe\footnote{With use of an unmotivated ``break-time''
model of AGs, Zeh, Klose and Hartmann (2003) have recently reached 
and standardized this old conclusion of ours.}.
In several more recent cases we used the CB model
to {\it predict} how the associated SN would compare with the rest of the AG
(GRB 011121, 020405, 021211; Dado et al.~2002b, 2002c, 2003d).
The most interesting case is that of
GRB 030329/SN2003dh, for which we also foretold the date when
the SN would be dominant enough to be discovered (Dado et al.~2003c). 

The fraction of visible GRBs, relative to the SNe that produce them,
is $f\sim 2\,\theta_{max}^2/(4\,\pi)$, where $\theta_{max}$ is
some effective value of the maximum observer's angle 
for which GRBs have been detectable. In practice $\theta_{max}$
is a few milliradians (Dado, Dar \& De R\'ujula 2002a).  This brings the 
{\it total}
rate of (observed or unobserved) GRBs to a value close to
that of core collapse SNe. This implies that a good fraction
of such SNe may produce long-duration GRBs\footnote{A
conclusion recently standardized by Lamb, Donaghy \& Graziani (2003).}
(Dar \& Plaga 1999; Dar \& De R\'ujula 2000).

\subsection{The GRBs and XRFs themselves}

In Dar \& De R\'ujula (2000) the initial Lorentz factor of the CBs was
argued to be $\gamma_0\sim 10^3$, a choice corroborated by all our
subsequent analyses. The high-energy photons of a single pulse in a GRB or
an XRF are produced as a CB coasts through the ``ambient light''
permeating the surroundings of the parent SN. The electrons enclosed in
the CB Compton up-scatter photons\footnote{Inverse Compton scattering by
narrow jets was proposed as {\it the mechanism} for GRB generation by Dar
\& Shaviv (1995). Much later, it was attributed in a first electronic
version of Lazzati et al.~(2003) to Lazzati et al.~(2000), an attribution
subsequently retracted (Lazzati et al.~2000, second version).  More
recently and apparently unaffected, Lazzati (2003) attributes the
mechanism once again to Lazzati et al.~(2000).} to energies that, close to
the CBs' direction of motion, correspond to the $\gamma$-rays of a GRB and
less close to it, to the X-rays of an XRF\footnote{This conclusion has
been enthusiastically standardized by Donaghy, Lamb, Graziani (2003) and
Lamb, Donaghy \& Graziani (2003a,b,c). Their model requires parent stars
that span faster in the distant past.}. Each pulse of a GRB or an XRF
corresponds to one CB. The timing sequence of emission of the successive
individual pulses (or CBs) reflects the chaotic accretion process and its
properties are not predictable, but those of the single pulses are (Dar \&
De R\'ujula 2003a and references therein). In practice GRBs are observable
for $\theta={\cal{O}}(1/\gamma_0)$, XRFs are the same phenomenon, observed
from somewhat larger angles 
(Dar \& De R\'ujula 2003a; Dado, Dar \& De R\'ujula 2003f).

GRBs are notorious for their variety. Yet, they have some two dozen common
properties. The characteristic $\gamma$-ray energy is surprisingly
narrowly distributed around $E=250$ keV. The energy spectra are well fit
by a specific ``Band" function.  The total ``equivalent spherical
energies'' are distributed over a wide range, that becomes very narrow for
the ``true'' (beaming-corrected) energy values. The photons arrive in
successive pulses with fairly similar time-profiles which, in ``long"
GRBs, have a median width of 0.5 seconds. The $\gamma$-ray polarization is
nearly maximal. A long list of phenomenologically-observed correlations
between many different GRB observables appear to be well satisfied. {\it
All} of these properties are very simply and successfully explained by the
CB model, in a manner that does not require the introduction of {\it any}
ad-hoc parameters (Dar \& De R\'ujula 2003a and references therein).

\subsection{GRB (and XRF) afterglows}

In the AG phase, and within a few observer minutes from the end of a GRB's
intense $\gamma$ emission, the emissivity of a CB is dominated by
synchrotron emission from the electrons that penetrate in it as it
propagates in the ISM. Integrated over frequency, this synchrotron
emissivity is proportional to the energy-deposition rate of the ISM
electrons in the CB. These electrons are ``Fermi-accelerated" in the CB's
tangled magnetic maze to a broken power-law energy distribution with a
``bend'' energy equal to their incident energy in the CBs' rest frame,
$E=\gamma(t)\,m_e\,c^2$. Their synchrotron radiation ---the afterglow---
is collimated and Doppler-boosted by the relativistic motion of the CBs.  
The radiation is also redshifted by the cosmological expansion and
attenuated on its way to an earthly observer, during its passage through
the CB itself, the host galaxy, the intergalactic space and our own
galaxy.

The approximate CB-model analysis of the fluences of GRB AGs, as functions of
time and frequency, is fairly simple. Besides $z$ and $\theta$ (which are
not parameters specific to the model) it involves only three quantities:
$\gamma_0$, $x_\infty$, an overall normalization, and $\nu_a$ (a {\it single}
free-parameter frequency pertinent to the radio-absorption within a CB).
In spite of its economy of parameters (notorious in the case of the
wide-band spectral shapes) the CB model satisfactorily describes ---in a
unified manner--- all measured GRB AGs, including that of GRB 980425 
(e.g., Dado et al.~2002a, 2003a, 2003e). Our impression is that no such 
claims can be made in the realm of standard GRB models.

\clearpage
\newpage

\clearpage
\newpage

\clearpage
\newpage

\clearpage
\newpage

\clearpage
\newpage

\clearpage
\newpage

\clearpage
\newpage

\clearpage
\newpage

\clearpage
\newpage

\end{document}